\documentclass[useAMS,usenatbib,usegraphicx,referee]{mn2e}

\usepackage{url}
\usepackage[dvips]{color} 

\newcommand{\Ni}{$^{56}$Ni}
\newcommand{\Co}{$^{56}$Co}

\newcommand{\ej}{\mathrm{ej}}
\newcommand{\csm}{\mathrm{csm}}
\newcommand{\s}{\mathrm{sh}}
\newcommand{\w}{\mathrm{w}}

\title[An Analytic Bolometric LC Model for Interaction-Powered SNe]
{
An Analytic Bolometric Light Curve Model of Interaction-Powered Supernovae
and its Application to Type IIn Supernovae
}
\author[T. J. Moriya et al.]
{Takashi J. Moriya$^{1,2,3}$\thanks{takashi.moriya@ipmu.jp},
Keiichi Maeda$^1$,
Francesco Taddia$^4$,
Jesper Sollerman$^4$, \newauthor
Sergei I. Blinnikov$^{5,6,7}$, and
Elena I. Sorokina$^7$
 \\
$^{1}$
Kavli Institute for the Physics and Mathematics of the Universe (WPI),
Todai Institutes for Advanced Study,
University of Tokyo, \\ 5-1-5 Kashiwanoha, Kashiwa, Chiba 277-8583, Japan
\\
$^{2}$
Department of Astronomy, Graduate School of Science, University of
Tokyo, 7-3-1 Hongo, Bunkyo, Tokyo 113-0033, Japan \\
$^{3}$
Research Center for the Early Universe, Graduate School of Science, University of Tokyo,\\
7-3-1 Hongo, Bunkyo, Tokyo 113-0033, Japan\\
$^{4}$
The Oskar Klein Centre, Department of Astronomy, Stockholm University,
AlbaNova, 10691 Stockholm, Sweden \\
$^{5}$
Institute for Theoretical and Experimental Physics, Bolshaya Cheremushkinskaya 25, 117218 Moscow, Russia\\
$^{6}$
Novosibirsk State University, Novosibirsk 630090, Russia\\
$^{7}$
Sternberg Astronomical Institute, M.V.Lomonosov Moscow State University,
Universitetski pr. 13, 119992 Moscow, Russia
}
\begin{document}

\date{Accepted 2013 July 24. Received 2013 July 24; in original form 2013 March 24}

\pagerange{\pageref{firstpage}--\pageref{lastpage}} \pubyear{2013}

\maketitle

\label{firstpage}

\begin{abstract}
We present an analytic model for bolometric light curves which are powered by the interaction between
supernova ejecta and a dense circumstellar medium.
This model is aimed at modeling Type IIn supernovae
to determine the properties of their supernova ejecta and
circumstellar medium.
Our model is not restricted to the case of
steady mass loss and can be applied broadly.
We only consider the case in which the optical depth of the unshocked
circumstellar medium is not high enough to affect the light curves.
We derive the luminosity evolution 
based on an analytic solution for the evolution of a dense shell
created by the interaction.
We compare our model bolometric light curves to observed bolometric
light curves of three Type IIn supernovae (2005ip, 2006jd, 2010jl)
and show that our model can constrain their
supernova ejecta and circumstellar medium properties.
Our analytic model is supported by numerical
light curves from the same initial conditions.
\end{abstract}


\section{Introduction}\label{sec1}
Massive stars which die as supernovae (SNe)
do not end their lives as they were born.
They change their mass, size, temperature, luminosity and many other
properties during their evolution toward their death.
The final fate, or the SN type, of a massive star is determined by
these changes during their evolution \citep[e.g.,][]{heger2003}.

One of the most critical factors which dramatically change the properties
of a star and largely affect its final fate is mass loss.
Massive stars continue to lose mass until the end of their lives
because of their huge luminosities.
From X-ray and radio observations of young SNe,
it has been possible to estimate the mass-loss rates
of SN progenitors shortly before their explosions
\citep[e.g.,][]{chevalier2006b,chevalier2006a,maeda2013}.
In most cases, the estimated mass-loss rates
are within the range expected from
the radiation-driven mass loss \citep[e.g.,][]{owocki2004}.

However, there are a number of SNe which seem to be strongly affected by
the interaction with 
circumstellar media (CSM) whose densities are too high to be explained
by the standard radiation-driven mass loss.
Most of them are spectroscopically classified as Type IIn
because of the narrow hydrogen emission lines seen in their spectra
\citep{schlegel1990,filippenko1997}.
The mass-loss rates of the progenitors before their
explosions are estimated to be $\sim10^{-4}-\sim0.1~M_\odot~\mathrm{yr^{-1}}$
\citep[e.g.,][]{taddia2013,kiewe2012,fox2011}.
Because of the high-density CSM, a cool dense shell is suggested
to be created between SN ejecta and CSM \citep[e.g.,][]{chevalier1994,chugai2004}.
The cool dense shell can create dust grains efficiently
and SNe IIn are promising sites for dust formation
\citep[e.g.,][]{kozasa2009}.
SNe IIn can also be used as a distance ladder thanks to the dense shell
\citep[e.g.,][]{blinnikov2012,potashov2012}.
In addition, some SNe IIn can be observed at very high redshifts
and may provide us with information
about star formation and initial-mass functions in the early Universe
\citep[e.g.,][]{cooke2008,cooke2009,cooke2012,tanaka2012,tanaka2013,whalen2013}.

Despite the many interesting phenomena associated with SNe IIn,
the nature of SNe IIn is still not well-understood.
Little is known about their progenitors or the
mechanisms that cause such extreme mass loss just before
the explosions.
There are several possible SN progenitors suggested to
cause extreme mass loss before the explosions,
e.g., super-asymptotic-giant-branch (AGB)
stars \citep[e.g.,][]{poelarends2008,botticella2009}
and massive red supergiants (RSGs) \citep[e.g.,][]{smith2009b,yoon2010}.
The progenitor of Type IIn SN 2008S which was found in archival images
is actually consistent with a super-AGB star \citep{prieto2008}.
The small degree of association between SN IIn sites and H$\alpha$
emitting regions within
their host galaxies also seem consistent with these relatively low-mass progenitors
(\citealt{anderson2012}, but see also \citealt{crowther2013}).
However, the other SN IIn progenitors detected in archival images
are very massive stars which are rather consistent with
luminous blue variables (LBVs)
(e.g.,
\citealt{gal-yam2009,mauerhan2012,pastorello2012},
see also \citealt{smith2011b,smith2011}).
LBVs are theoretically interpreted as an evolutionary stage 
to a Wolf-Rayet star and they have not been considered
as a pre-SN phase (e.g., \citealt{crowther2007,maeder2000}, but see \citealt{groh2013,moriya2013c,langer2012}).
In addition, extreme mass-loss mechanisms of such very massive stars
are not well-known.
There are several suggested mechanisms to induce extreme
mass loss, like
pulsational pair-instability (e.g., \citealt{woosley2007}),
rotation (e.g., \citealt{maeder2001}),
porosity (e.g., \citealt{owocki2004}),
gravity-mode oscillations (e.g., \citealt{quataert2012}), or
binary interaction (e.g., \citealt{chevalier2012}),
but we still do not know which mechanisms are actually
related to SNe IIn (see also \citealt{dwarkadas2011}).

For a better understanding of SNe IIn,
especially their progenitors and extensive mass-loss mechanisms,
we need more theoretical
investigation of SNe IIn as well as more observational data.
In this paper we develop a simple analytic bolometric LC model
which can be used to fit SN IIn observations
to estimate CSM and SN ejecta properties.
We believe that the information obtained
by applying our LC model to many SNe IIn
can lead to a better understanding of SNe IIn.

This paper is organized as follows.
We present our analytic LC model in Section \ref{sec2}.
We derive the evolution of the shocked shell analytically and 
use it to obtain the bolometric LC evolution.
We apply it to some observational SN IIn bolometric LCs 
and obtain constraints on the properties of SN ejecta and CSM for these SNe in Section \ref{sec3}.
The discussion is given in Section \ref{sec4} and
we conclude this paper in Section \ref{sec5}.

\section{Analytic Bolometric Light Curve Model}\label{sec2}
In this section, we develop an analytic SN bolometric LC model 
under the assumption that its main power source is the kinetic
energy of SN ejecta colliding with a dense CSM.
At first, we analytically investigate the evolution of the dense
shell created by the interaction in Section \ref{evlofshell}.
The analytic solution for the evolution of the dense shell
before time $t_t$ (see below)
is essentially the same as obtained
in previous works \citep[e.g.,][]{chevalier1982a,chevalier1990,
chevalier1994,chevalier2003} but
our solution does not assume steady mass loss
(see also \citealt{fransson1996}).

After deriving the evolution of the dense shell, we provide an analytic
expression for bolometric LCs.
This method was introduced by \citet{chugai1994}
\citep[see also][]{wood-vasey2004,svirski2012}
to explain the luminoisity due to the interaction 
but their model assumes a CSM from steady mass loss.
We generalize this method for the cases of non-steady mass loss
and apply our model to entire bolometric LCs.
\citet{manos2012,manos2013} also follow a similar approach to obtain an analytic
LC model from the interaction but they consider the case where
the unshocked CSM is optically thick.
Here, we consider the case in which the unshocked CSM does not affect the
bolometric LC so much.
Some SNe IIn are suggested to have very optically thick CSM
to explain their huge luminosities \citep[e.g.,][]{chevalier2011,moriya2012,moriya2013b,ginzburg2012}
but they are beyond the scope of this paper.
High energy photons are expected to be emitted when the CSM is optically
thin \citep[e.g.,][]{chevalier1994}
but they are presumed to be absorbed by the dense shell because of its
high column density and re-emitted as optical photons
which are mainly observed \citep[e.g.,][]{wilms2000}.
The inverse Compton scattering and other effects can also reduce the energy of
the photons \citep[e.g.,][]{chevalier2012b}.

\subsection{Evolution of the Shocked Dense Shell}\label{evlofshell}
\subsubsection{General Case}\label{evlofshellgeneral}
The shocked dense CSM and SN ejecta form a thin dense
shell because of the efficient radiative cooling.
We assume that the thickness of the shocked shell is much smaller than
its radius and it can be denoted by a radius $r_\s(t)$.
The conservation of momentum requires
\begin{equation}
M_\s\frac{dv_\s}{dt}=4\pi r_\s^2\left[\rho_\ej (v_\ej-v_\s)^2-\rho_\csm (v_\s-v_\w)^2 \right],
\end{equation}
where $M_\s$ is the total mass of the shocked SN ejecta and CSM,
$v_\s$ is the velocity of the shell, $\rho_\ej$ is the SN ejecta density,
$v_\ej$ is the SN ejecta velocity, $\rho_\csm$ is the CSM density, and
$v_\w$ is the CSM velocity.
We derive the evolution of $r_\s$ based on this equation.
We do not use the equation for the conservation of energy
which is necessary to derive the self-similar solution
including the reverse shock and the forward shock \citep{chevalier1982b,nadyozhin1985}.
This is because of the radiative energy loss from the dense shell.
When the radiative cooling is efficient,
the shocked region does not extend as wide as the width 
expected from the self-similar solution due to 
the loss of the thermal pressure caused by the radiative energy loss.
Thus, our approximation to neglect the shell width
is presumed to be a good approximation to the shell evolution.

We further assume that the CSM density follows $\rho_\csm =D r^{-s}$ and
that the CSM velocity $v_\w$ is constant.
We adopt a double power-law profile
for the density of homologously $(v_\ej=r/t)$
expanding SN ejecta
($\rho_\ej\propto r^{-n}$ outside and $\rho_\ej\propto r^{-\delta}$ inside)
based on numerical simulations of SN explosions \citep[e.g.,][]{matzner1999}.
With SN kinetic energy $E_\ej$ and
SN ejecta mass $M_\ej$, the SN density structure is
expressed as
\begin{equation}
\rho_\ej\left(v_\ej,t\right)=\left\{ \begin{array}{ll}
\frac{1}{4\pi(n-\delta)}
\frac{[2(5-\delta)(n-5)E_\ej]^{(n-3)/2}}{
[(3-\delta)(n-3)M_\ej]^{(n-5)/2}}
t^{-3}v_\ej^{-n} & (v_\ej>v_t), \\ 
\frac{1}{4\pi(n-\delta)}
\frac{[2(5-\delta)(n-5)E_\ej]^{(\delta-3)/2}}{
[(3-\delta)(n-3)M_\ej]^{(\delta-5)/2}}
t^{-3}v_\ej^{-\delta} &
 (v_\ej<v_t), \\ 
\end{array} \right.
\end{equation}
where $v_t$ is obtained from the density continuity condition at the interface
of the two density structures as well as $E_\ej$ and $M_\ej$ as follows,
\begin{equation}
v_t=\left[
\frac{2(5-\delta)(n-5)E_\ej}{(3-\delta)(n-3)M_\ej}
\right]^{\frac{1}{2}}.
\end{equation}

The outer density slope $n$ depends on the SN progenitor and
$n\simeq 7$ ($n=6.67$ exactly) is the lowest possible $n$ expected from the self-similar
solution of \citet{sakurai1960} \citep[e.g.,][]{chevalier1990}.
A value of $n\simeq 10$ is expected for SN Ib/Ic and SN Ia progenitors \citep{matzner1999,kasen2010}
and $n\simeq 12$ is expected for explosions of RSGs \citep{matzner1999}. 
The inner density slope $\delta$ is $\delta\simeq 0-1$.

At first, the outer SN ejecta with $\rho_\ej\propto r^{-n}$ starts to interact
with the CSM. In this phase, $M_\s$ becomes
\begin{eqnarray}
M_\s &=& \int^{r_\s}_{R_\mathrm{p}} 4\pi r^2\rho_\csm dr
+ \int_{v_\ej/t}^{v_\mathrm{ej,max}/t}4\pi r^2 \rho_\ej dr, \label{s=2mass1}\\
&=& \frac{4\pi D}{3-s}r_\s^{3-s}+
\frac{t^{n-3}}{(n-\delta)(n-3)r_\s^{n-3}}
\frac{[2(5-\delta)(n-5)E_\ej]^{(n-3)/2}}{
[(3-\delta)(n-3)M_\ej]^{(n-5)/2}},\label{s=2mass2}
\end{eqnarray}
where $R_\mathrm{p}$ is the radius
of the progenitor, $v_\mathrm{ej,max}$ is the velocity of the outermost
layer of the SN ejecta before the interaction.
In deriving Equation (\ref{s=2mass2}) from Equation (\ref{s=2mass1}),
we have assumed $r_\s\gg R_\mathrm{p}$, $v_\mathrm{ej,max}\gg v_\ej$, and $s<3$.

With the above equations and $v_\ej=r_\s/t$ (homologous expansion of the SN ejecta),
the equation for the conservation of momentum becomes
\begin{eqnarray}
\left[
\frac{4\pi D}{3-s}r_\s^{3-s}+
\frac{t^{n-3}}{(n-\delta)(n-3)r_\s^{n-3}}
\frac{[2(5-\delta)(n-5)E_\ej]^{(n-3)/2}}{
[(3-\delta)(n-3)M_\ej]^{(n-5)/2}}
\right]
\frac{d^2r_\s}{dt^2} \nonumber
= \ \ \ \ \ \ \ \ \ \ \ \ \ \ \ \ \ \ \ \ \ \ \ \ \\ 
\frac{1}{(n-\delta)}
\frac{[2(5-\delta)(n-5)E_\ej]^{(n-3)/2}t^{n-3}}{
[(3-\delta)(n-3)M_\ej]^{(n-5)/2}r_\s^{n-2}}
\left(\frac{r_\s}{t}-\frac{d r_\s}{dt}\right)^2-4\pi D r_\s^{2-s}
\left(\frac{dr_\s}{dt}\right)^2.
\end{eqnarray}
Here, we assume that the CSM velocity is much smaller than the shell
velocity $(v_\s\gg v_\w)$.
Solving the differential equation, we get a power-law solution
\begin{equation}
r_\s (t)=
\left[
\frac{(3-s)(4-s)}{4\pi D(n-4)(n-3)(n-\delta)}
\frac{[2(5-\delta)(n-5)E_\ej]^{(n-3)/2}}{[(3-\delta)(n-3)M_\ej]^{(n-5)/2}}
\right]^{\frac{1}{n-s}}t^{\frac{n-3}{n-s}}. \label{rshbefore}
\end{equation}
Note that $r_\s$ obtained with this approach has the same time dependence
as the self-similar solution $[t^{(n-3)/(n-s)}]$
\citep{chevalier1982a,chevalier1982b,chevalier2003,nadyozhin1985}.

Equation (\ref{rshbefore}) holds until the time $t_t$ when
the interacting region reaches down to the inner ejecta,
namely when the $v_\ej$ entering the shell becomes $v_t$ or
$r_\s (t_t)= v_t t_t$ is satisfied,
i.e.,
\begin{equation}
t_t=\left[\frac{(3-s)(4-s)}{4\pi D(n-4)(n-3)(n-\delta)}
\frac{[(3-\delta)(n-3)M_\ej]^{(5-s)/2}}{[2(5-\delta)(n-5)E_\ej]^{(3-s)/2}}
\right]^{\frac{1}{3-s}}.
\end{equation}
After $t_t$, the density structure of the SN ejecta entering the shell
starts to follow $\rho_\ej\propto r^{-\delta}$ and the equation of
the momentum conservation becomes
\begin{eqnarray}
\left[
\frac{4\pi D}{3-s}r_\s^{3-s}+M_\ej
-\frac{r_\s^{3-\delta}}{(n-\delta)(3-\delta)t^{3-\delta}}
\frac{[2(5-\delta)(n-5)E_\ej]^{(\delta-3)/2}}{
[(3-\delta)(n-3)M_\ej]^{(\delta-5)/2}}
\right]
\frac{d^2r_\s}{dt^2} \nonumber
=  \ \ \ \ \ \ \ \ \ \ \ \ \ \ \ \ \ \\ 
\frac{1}{(n-\delta)}
\frac{[2(5-\delta)(n-5)E_\ej]^{(\delta-3)/2}r_\s^{2-\delta}}{
[(3-\delta)(n-3)M_\ej]^{(\delta-5)/2}t^{3-\delta}}
\left(\frac{r_\s}{t}-\frac{d r_\s}{dt}\right)^2-4\pi D r_\s^{2-s}
\left(\frac{dr_\s}{dt}\right)^2.\label{eqafter}
\end{eqnarray}
Generally, we cannot solve Equation (\ref{eqafter}) analytically but
the solution of Equation (\ref{eqafter})
is expected to asymptotically approach
the solution of the differential equations
\begin{eqnarray}
M_\s\frac{d^2r_\s}{dt^2}&=&4\pi r_\s^2 (-\rho_\csm
 v_\s^2), \\
\left(\frac{4\pi D}{3-s}r_\s^{3-s}+M_\ej\right)\frac{d^2r_\s}{dt^2}&=&
-4\pi Dr_\s^{2-s}\left(\frac{dr_\s}{dt}\right)^2.\label{asympeq}
\end{eqnarray}
The asymptotic solution from Equation (\ref{asympeq}) satisfies the equation
\begin{equation}
\frac{4\pi D}{4-s}r_\s(t)^{4-s}+(3-s)M_\ej r_\s(t)-(3-s)M_\ej
\left(\frac{2E_\ej}{M_\ej}\right)^{\frac{1}{2}}t=0.
\label{longafter}
\end{equation}
The boundary conditions
\begin{eqnarray}
r_\s(t=0)&=&0, \\
\frac{dr_\s}{dt}(t=0)&=&\left(\frac{2 E_\ej}{M_\ej}\right)^{\frac{1}{2}},
\end{eqnarray}
are applied in deriving Equation (\ref{longafter}).
As the asymptotic solution is derived by assuming that most of the
SN ejecta is in the dense shell, the dependence of $r_\s (t)$
on the SN ejecta structure ($n$ and $\delta$) disappears.

\subsubsection{Case of Steady Mass Loss (s=2)}
Here, we write down $r_\s (t)$ derived in the previous section
for the special case of the steady mass loss $(s=2)$.
The CSM density structure becomes
\begin{equation}
\dot{M}=4\pi r^2 \rho_\csm v_\w,
\end{equation}
where $\dot{M}$ is the mass-loss rate
and $D$ can be expressed as
\begin{equation}
D=\frac{\dot{M}}{4\pi v_\w}.
\end{equation}
Then, $r_\s(t)$ before $t=t_t$ is
\begin{equation}
r_\s (t)=
\left[
\frac{2}{(n-4)(n-3)(n-\delta)}
\frac{[2(5-\delta)(n-5)E_\ej]^{(n-3)/2}}{[(3-\delta)(n-3)M_\ej]^{(n-5)/2}}
\frac{v_\w}{\dot{M}}
\right]^{\frac{1}{n-2}}t^{\frac{n-3}{n-2}}
\ \ (t<t_t), \label{rshbefore-s=2}
\end{equation}
and
\begin{equation}
t_t=\frac{2}{(n-4)(n-3)(n-\delta)}
\frac{[(3-\delta)(n-3)M_\ej]^{3/2}}{[2(5-\delta)(n-5)E_\ej]^{1/2}}\frac{v_\w}{\dot{M}}.
\end{equation}
The asymptotic solution after $t_t$ becomes
\begin{equation}
r_\s(t)=\frac{v_\w}{\dot{M}}M_\ej\left[-1+
\left(1+2\sqrt{\frac{2 E_\ej}{M_\ej^3}}\frac{\dot{M}}{v_\w}t\right)^{\frac{1}{2}}
\right].\label{longafter-s=2}
\end{equation}
As noted in Section \ref{evlofshellgeneral},
the asymptotic solution is independent of the SN density structure ($n$
and $\delta$).

\subsection{Bolometric Light Curve}
\subsubsection{General Case}
\begin{figure}
\begin{center}
 \includegraphics[width=0.75\columnwidth]{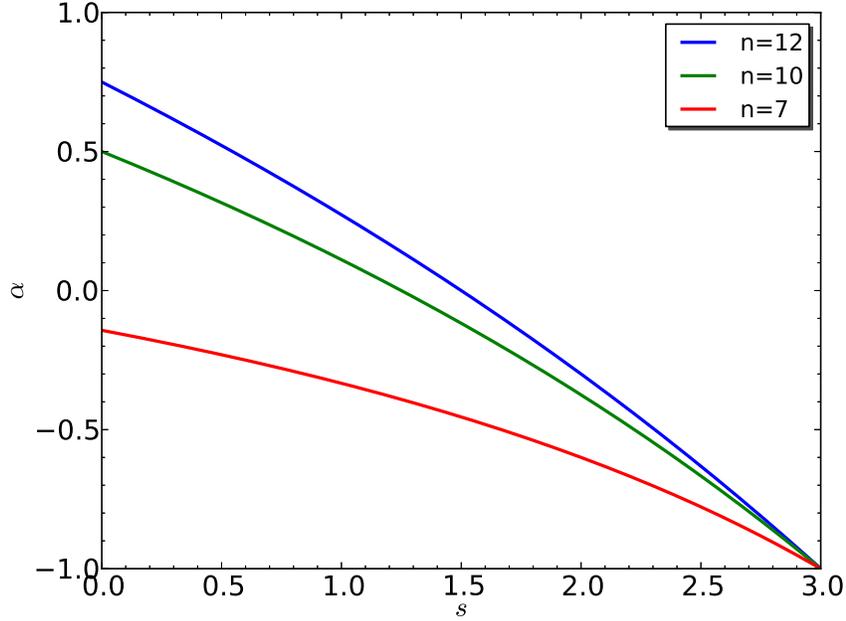}
  \caption{
$\alpha$ ($L\propto t^\alpha$ before $t_t$) as a function of $s$ for some $n$.
$n\simeq12$ represents RSG explosions and $n\simeq10$ is for SNe Ib/Ic and SNe Ia
progenitors. $n\simeq 7$ is the minimum possible $n$.
}
\label{alpha}
\end{center}
\end{figure}

We construct an analytic bolometric LC based on $r_\s(t)$
obtained in the previous section.
We assume that the kinetic energy of the SN ejecta is the dominant source
of the SN luminosity.
The available kinetic energy is
\begin{equation}
dE_\mathrm{kin} = 4\pi r_\s^2 \frac{1}{2} \rho_\csm v_\s^2 dr_\s,
\end{equation}
and thus the bolometric luminosity will be
\begin{equation}
L=\epsilon \frac{dE_\mathrm{kin}}{dt}
= 2\pi \epsilon \rho_\csm r_\s^2 v_\s^3,
\end{equation}
where $\epsilon$ is the conversion efficiency from kinetic energy to radiation.
Especially, the bolometric luminosity before $t_t$ can be expressed
as a power-law function
\begin{equation}
L=L_1 t^\alpha,
\end{equation}
where
\begin{equation}
L_1=
\frac{\epsilon}{2}\left(4\pi D\right)^{\frac{n-5}{n-s}}
\left(\frac{n-3}{n-s}\right)^3
\left[
\frac{(3-s)(4-s)}{(n-4)(n-3)(n-\delta)}
\frac{[2(5-\delta)(n-5)E_\ej]^{(n-3)/2}}{[(3-\delta)(n-3)M_\ej]^{(n-5)/2}}
\right]^{\frac{5-s}{n-s}},
\end{equation}
\begin{equation}
\alpha=\frac{6s-15+2n-ns}{n-s}.\label{alphaeq}
\end{equation}
In Figure \ref{alpha}, $\alpha$ is plotted as a function of $s$ for $n=12,10,7$.

After $t_t$, the asymptotic bolometric LC can be obtained based on the
asymptotic radius evolution from Equation (\ref{longafter}).
\begin{eqnarray}
L&=&2\pi\epsilon\rho_\csm r_\s^2v_\s^3 \\
&=&2\pi\epsilon
 Dr_\s^{2-s}\left[\frac{(3-s)M_\ej\left(\frac{2E_\ej}{M_\ej}\right)^{\frac{1}{2}}}{
4\pi Dr_\s^{3-s}+(3-s)M_\ej}\right]^3. \label{asymgen}
\end{eqnarray}

\subsubsection{Case of Steady Mass Loss (s=2)}
In the case of steady mass loss $(s=2)$, we can use $4\pi D=\dot{M}/v_\w$
and express $L$ before $t_t$ as
\begin{eqnarray}
L&=&\frac{\epsilon}{2}\frac{\dot{M}}{v_\w}v_\s^3 \\
&=&\frac{\epsilon}{2}\left(\frac{\dot{M}}{v_\w}\right)^{\frac{n-5}{n-2}}
\left(\frac{n-3}{n-2}\right)^3
\left[
\frac{2}{(n-4)(n-3)(n-\delta)}
\frac{[2(5-\delta)(n-5)E_\ej]^{(n-3)/2}}{[(3-\delta)(n-3)M_\ej]^{(n-5)/2}}
\right]^{\frac{3}{n-2}}t^{-\frac{3}{n-2}}. \ \ \ \ \ \ 
\end{eqnarray}
This equation is basically the same as obtained in previous
studies \citep[e.g.,][]{chugai1994,wood-vasey2004}.

We can also express the asymptotic bolometric LC after $t_t$ using Equation (\ref{longafter-s=2}).
\begin{equation}
L=\frac{\epsilon}{2}\frac{\dot{M}}{v_\w}
\left(\frac{2E_\ej}{M_\ej}\right)^{\frac{3}{2}}
\left[1+2\frac{\dot{M}}{v_w}
\left(\frac{2E_\ej}{M_\ej^3}\right)^{\frac{1}{2}}t\right]^{-\frac{3}{2}}.
\label{s=2longafter}
\end{equation}

By defining two parameters $a$ and $b$ as
\begin{eqnarray}
a&=&\frac{\epsilon}{2}\frac{\dot{M}}{v_\w}\left(\frac{2E_\ej}{M_\ej}\right)^{\frac{3}{2}}, \\
b&=&2\frac{\dot{M}}{v_\w}\left(\frac{2E_\ej}{M_\ej^3}\right)^{\frac{1}{2}},
\end{eqnarray}
we can express $L$ in a simple way. Namely,
\begin{equation}
L=2^{-\frac{3(n-7)}{2(n-2)}}ab^{-\frac{3}{n-2}}
\left(\frac{n-3}{n-2}\right)^3
\frac{\left[2(5-\delta)(n-5)\right]^{\frac{3(n-3)}{2(n-2)}}}{
\left[(n-4)(n-3)(n-\delta)\right]^{\frac{3}{n-2}}
\left[(3-\delta)(n-3)\right]^{\frac{3(n-5)}{2(n-2)}}}
t^{-\frac{3}{n-2}},\label{earlye}
\end{equation}
before $t_t$ and
\begin{equation}
L=a\left(1+bt\right)^{-\frac{3}{2}}, \label{latee}
\end{equation}
long after $t_t$. Here, $t_t$ is expressed as
\begin{equation}
t_t=\frac{4\left[(3-\delta)(n-3)\right]^{\frac{3}{2}}}{
(n-4)(n-3)(n-\delta)\left[(5-\delta)(n-5)\right]^{\frac{1}{2}}b}.
\end{equation}
The physical parameters of CSM and SN ejecta have the relations
\begin{eqnarray}
E_\ej&=&\frac{2a}{\epsilon b}, \label{Eej}\\
\frac{\dot{M}}{v_\w}M_\ej^{-\frac{3}{2}}&=&
\frac{1}{4}\left(\frac{\epsilon b^3}{a}\right)^{\frac{1}{2}}.\label{Mej}
\end{eqnarray}

\section{Comparison with Observations}\label{sec3}
\subsection{Procedures}
We first show examples of procedures to fit our analytic bolometric LC
to bolometric LCs constructed from observations.
The actual processes for the comparison depend on the available
information from observations but the basic concepts 
will be essentially the same as the examples presented here.

Our bolometric LC model consists of two components.
Before $t_t$, the model LC has a power-law dependence on time ($L=L_1 t^{\alpha}$).
Thus, we can first use the function $L_1t^\alpha$ to fit an early LC
and obtain $L_1$ and $\alpha$.
Assuming $n$, the CSM density slope $s$
can be constrained just by $\alpha$ through Equation (\ref{alphaeq})
(Figure \ref{alpha}).

If there are spectral observations at these epochs
and the shell velocity evolution can be estimated by them, we can use
\begin{eqnarray}
v_\s(t)&=&\frac{dr_\s}{dt}\\
&=&\frac{n-3}{n-s}
\left[\frac{(3-s)(4-s)}{4\pi D (n-4)(n-3)(n-\delta)}
\frac{[2(5-\delta)(n-5)E_\ej]^{(n-3)/2}}{[(3-\delta)(n-3)M_\ej]^{(n-5)/2}}
\right]^{\frac{1}{n-s}}t^{-\frac{3-s}{n-s}}\\
&\equiv&v_1 t^{-\frac{3-s}{n-s}},
\end{eqnarray}
to fit the velocity evolution and obtain $v_1$.
Just from the three values, $L_1$, $\alpha$, and $v_1$, we can obtain the
CSM density structure for given $\epsilon$ and $n$,
\begin{equation}
D=\frac{1}{2\pi\epsilon}\left(\frac{n-3}{n-s}\right)^{2-s}L_1v_1^{s-5}.
\label{generalD}
\end{equation}
This means that we can estimate the mass-loss rate without assuming $M_\ej$ and $E_\ej$.
As the time dependence of $v_\s$ is small, the velocity information
of just a single epoch can constrain $D$.
So far, $M_\ej$ and $E_\ej$ are degenerated and we have to assume either $M_\ej$
or $E_\ej$ to constrain the other parameter.

The formulae $L=L_1t^\alpha$ and $v_\s=v_1 t^{-(3-s)/(n-s)}$
can only be applied before $t_t$. After obtaining the physical values,
we have to check whether $t_t$ is larger than the epochs used for the fitting.
If there is an available bolometric LC after $t_t$,
we can fit Equation (\ref{asymgen}) to the LC and obtain further
constraints to break the degeneracy between $E_\ej$ and $M_\ej$.

We show how this procedure works in the next section by
using actual bolometric LCs from observations.

\subsection{Examples}\label{exmpls}
Here we compare our analytic bolometric LCs to observed LCs of
SNe IIn 2005ip, 2006jd, and 2010jl,
and estimate CSM and SN ejecta properties of them.
We assume $\epsilon=0.1$ in this section. 
$\epsilon$ is affected by SN ejecta mass and CSM mass but
it is typically of the order of 0.1 \citep[e.g.,][]{moriya2013b}.
All the fitting procedures in this section are performed by
using the least-squares method unless otherwise mentioned.
Table \ref{tabpropsum} is a summary of the SN IIn properties
derived in this section.

\begin{table*}
\centering
\begin{minipage}{80mm}
\caption{SN IIn properties estimated by the bolometric LC model $(\epsilon=0.1)$.}
\label{tabpropsum}
\begin{tabular}{ccccc}
\hline
SN&\multicolumn{2}{c}{$s$ ($\rho_\csm\propto r^{-s}$)}& $\langle\dot{M}\rangle^a$ &$E_\ej$ \\
& $n=10$ & $n=12$ &($M_\odot~\mathrm{yr^{-1}}$) &($10^{51}$ erg) \\
\hline
2005ip &2.3&2.4&$1.2-1.4\times 10^{-3}$&13-15$^b$ \\
2006jd &1.4&1.6&$1.3-1.7\times 10^{-3}$&12-13$^b$\\
2010jl &\multicolumn{2}{c}{2.2$^b$}&0.039$^{b}$&23$^{b}$\\
\hline
\end{tabular}
\footnotetext{$^a$Average rate derived by the CSM mass within $10^{16}$
 cm and the CSM wind velocity $100~\mathrm{km~s^{-1}}$.}
\footnotetext{$^b$Derived assuming $M_\ej=10~M_\odot$.}
\end{minipage}
\end{table*}

\subsubsection{SN 2005ip}\label{sec:05ip}
SN 2005ip was intensively observed by \citet{stritzinger2012}
from ultraviolet to near-infrared wavelengths. They
derived a bolometric LC that we use for the comparison
to our bolometric LC model.
Optical photometric and spectroscopic observations are
also reported by \citet{smith2009}, whereas
\citet{fox2009,fox2010,fox2011,fox2013} summarize the near-infrared observations
of SN 2005ip. 
We assume that the explosion date of SN 2005ip was 9 days before
its discovery and all the following dates are since the explosion.

\begin{figure}
\begin{center}
 \includegraphics[width=0.49\columnwidth]{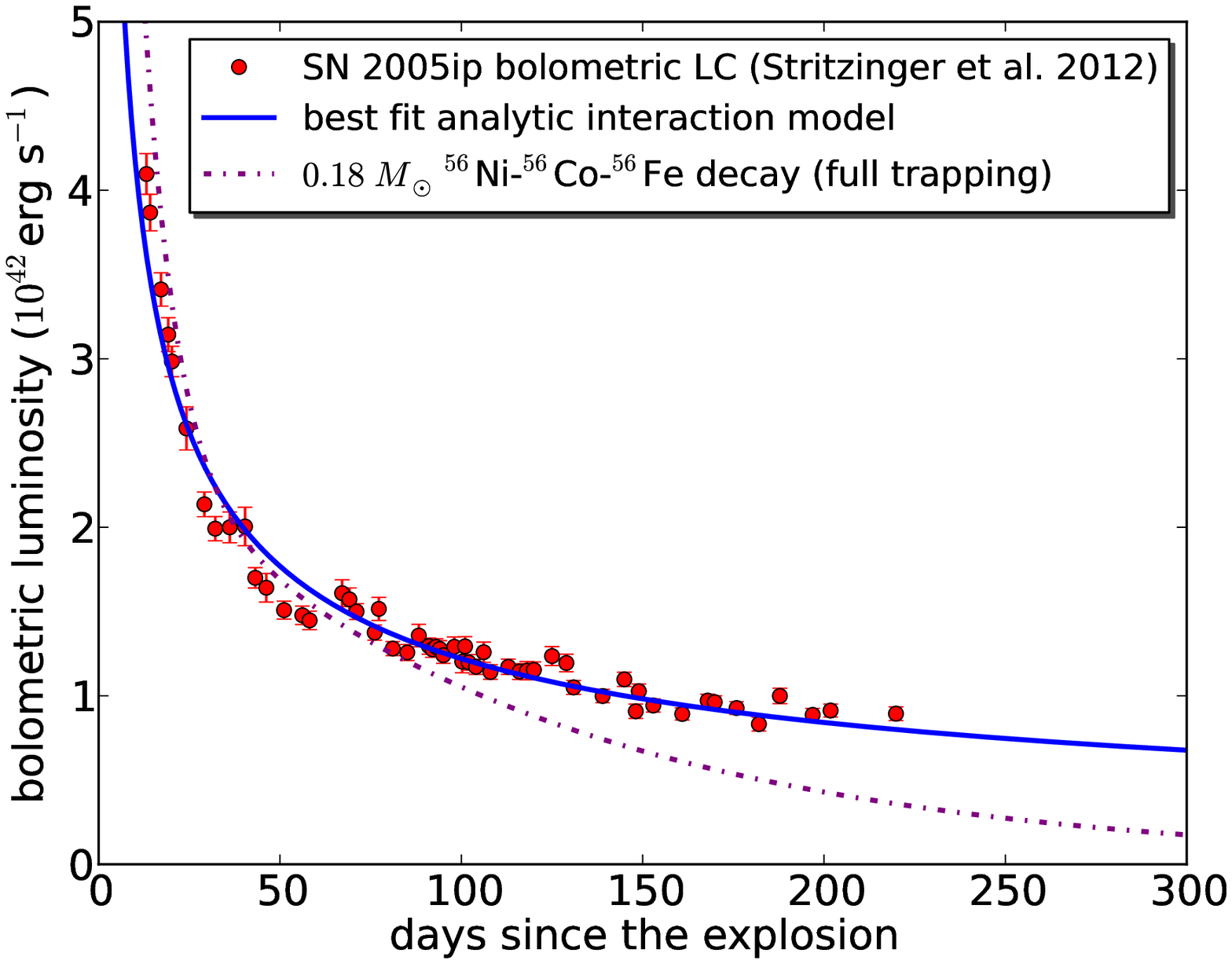}
 \includegraphics[width=0.49\columnwidth]{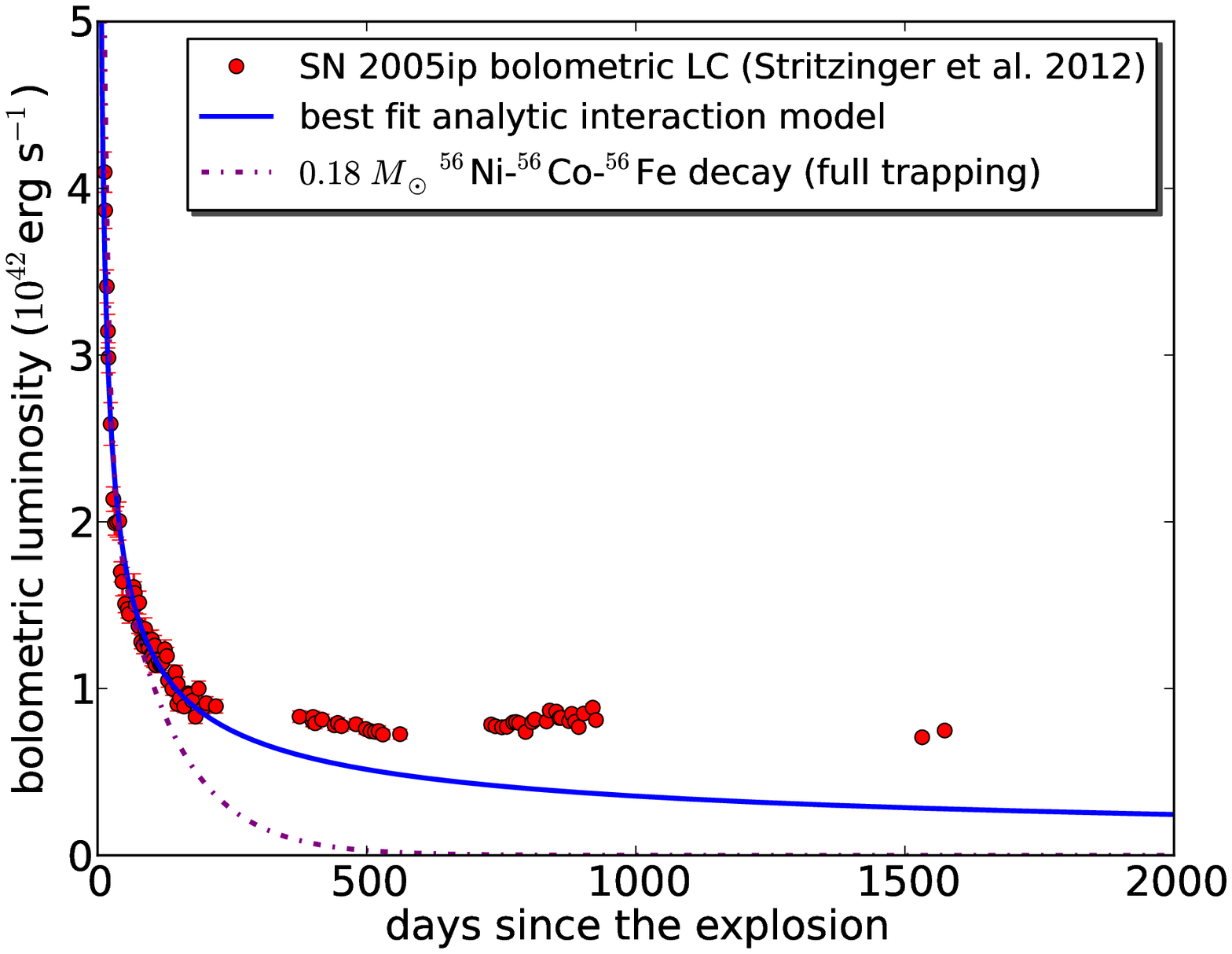}
  \caption{
Bolometric LC of SN 2005ip \citep{stritzinger2012} and some LC models.
The solid line is the best fit to $L=L_1t^\alpha$ up to 220 days.
The dot-dashed line is the available energy from the radioactive decay
 of $0.18~M_\odot$ \Ni.
The luminosity inputs from the two power souces resemble each other
up to about 100 days and the later LC is required to distinguish
between them. 
}
\label{sn2005ip}
\end{center}
\end{figure}

At first, we fit the obtained bolometric LC up to 220 days by
$L=L_1t^\alpha$ and we get
\begin{equation}
L=1.44\times 10^{43}\left(\frac{t}{\mathrm{1~day}}\right)^{-0.536}
\ \mathrm{erg~s^{-1}}.
\end{equation}
In Figure \ref{sn2005ip} we show the result.
$\alpha=-0.536$ corresponds to $s=2.3$ and $s=2.4$ for $n=10$ and
$n=12$, respectively.
Thus, the CSM around the progenitor of SN 2005ip likely had slightly
steeper density structure than the expected density structure from
steady mass loss.
The deviation from the steady mass loss of SN IIn progenitors
is also suggested from X-ray observations \citep{dwarkadas2012}.

One interesting feature in Figure \ref{sn2005ip}
is the similarity of the analytic LC from the SN ejecta-CSM interaction
to the available energy from the radioactive decay of $0.18~M_\odot$ \Ni\
before around 100 days. The available energy
from the radioactive decay
$^{56}\mathrm{Ni}\rightarrow$$^{56}\mathrm{Co}\rightarrow$$^{56}\mathrm{Fe}$
is \citep{nadyozhin1994}
\begin{equation}
\left[6.45\times 10^{43}e^{-t/(8.8~\mathrm{days})}+1.45\times 10^{43}e^{-t/(111.3~\mathrm{days})}\right]
\frac{M_{^{56}\mathrm{Ni}}}{M_\odot}~\mathrm{erg~s^{-1}}, \label{radioactive}
\end{equation}
where $M_{^{56}\mathrm{Ni}}$ is the initial \Ni\ mass.
We cannot distinguish between the two power sources 
only from the bolometric LC before about 100 days.
The two energy sources can only be distinguished by LCs at later epochs.
The similarity, especially at around 50 days, is because of the decay time of \Co.
At around 50 days, the radioactive energy from \Co\ is dominant and the
available energy from the decay follows $\propto e^{-t/(111.3~\mathrm{days})}$.
The values and the decline rates (the first derivatives)
of the functions following $\propto e^{-t/(111.3~\mathrm{days})}$ and
$\propto t^{-m}$ ($m$ is a constant) can get similar
at $t=111.3m~\mathrm{days}$. Looking at Figure \ref{alpha},
$m\simeq 0.5$ at around $s\simeq2$, so the two functions can be
similar at around $t\simeq 50$ days.
For a LC from the interaction to have a similar decline rate
to that from the \Co\ radioactive decay after $\simeq 100$ days, $m$ should be close to unity and the CSM density slope should be
steep $(s\simeq 3)$.

The shell velocity of SN 2005ip around 100 days since the explosion is likely
$\simeq 17,500~\mathrm{km~s^{-1}}$ \citep{stritzinger2012}.
Then, based on Equation (\ref{generalD}), we get
\begin{equation}
\rho_\csm(r)=\left\{ \begin{array}{ll}
8.4\times 10^{-16}\left(\frac{r}{10^{15}~\mathrm{cm}}\right)^{-2.3}~\mathrm{g~cm^{-3}} & (n=10), \\ 
1.0\times 10^{-15}\left(\frac{r}{10^{15}~\mathrm{cm}}\right)^{-2.4}~\mathrm{g~cm^{-3}} & (n=12). \\ 
\end{array} \right.
\end{equation}
The Thomson scattering optical depth $\tau_\s$ of the solar-metallicity
unshocked CSM when the shell is at the radius $10^{15}$ cm,
above which the shell is located at the epochs we fit the LC, is
\begin{equation}
\tau_\s=\left\{ \begin{array}{ll}
0.22 & (n=10), \\ 
0.25 & (n=12), \\ 
\end{array} \right.
\end{equation}
and our assumption that the unshocked CSM does not affect the LC
at the epochs we use for the fitting is justified.
We estimate an average mass-loss rate $\langle \dot{M}\rangle$
by using the CSM mass within $10^{16}$ cm.
Assuming $v_\w=100~\mathrm{km~s^{-1}}$, the CSM mass within $10^{16}$ cm
is lost from the progenitor
in 32 years before the explosion. The average mass-loss
rate in this period is
\begin{equation}
\langle \dot{M}\rangle=\left\{ \begin{array}{ll}
1.2\times 10^{-3}~M_\odot~\mathrm{yr^{-1}} & (n=10), \\ 
1.4\times 10^{-3}~M_\odot~\mathrm{yr^{-1}} & (n=12). \\ 
\end{array} \right.\label{05iprate}
\end{equation}

The bolometric luminosity of SN 2005ip after 300 days is almost constant
($\simeq 8.2\times 10^{41}~\mathrm{erg~s^{-1}}$).
The asymptotic solution (Equation \ref{asymgen}) can have a constant luminosity at
a certain condition. For example, the asymptotic solution
for $s=2$ (Equation \ref{latee}) can be constant if $bt\ll 1$.
However, for the case of SN 2005ip,
we could not find a constant asymptotic solution which is consistent
with the early LC before 300 days.
The constant luminosity may be due to, e.g., another CSM component or
light echos.

To constrain the SN properties, we assume $M_\ej=10~M_\odot$.
Then, from $L_1$ above, we obtain
\begin{equation}
E_\ej=\left\{ \begin{array}{ll}
1.3\times 10^{52}~\mathrm{erg~s^{-1}} & (n=10), \\ 
1.5\times 10^{52}~\mathrm{erg~s^{-1}} & (n=12). \\ 
\end{array} \right.
\end{equation}
$t_t$ becomes
\begin{equation}
t_t=\left\{ \begin{array}{ll}
4.2\times 10^3~\mathrm{days} & (n=10), \\ 
5.0\times 10^3~\mathrm{days} & (n=12), \\ 
\end{array} \right.
\end{equation}
so the epochs we used for the fitting $(t<220~\mathrm{days})$ are justified.

The average mass-loss rate we obtained ($10^{-3}~M_\odot~\mathrm{yr^{-1}}$) 
is consistent with the rate estimated by \citet{fox2011}
($1.8\times 10^{-3}~M_\odot~\mathrm{yr^{-1}}$) but
larger than the rate suggested by \citet{smith2009}
($2\times 10^{-4}~M_\odot~\mathrm{yr^{-1}}$).
Based on these mass-loss rates, \citet{smith2009} conclude that
the progenitor of SN 2005ip is a massive RSG like VY CMa
\citep[e.g.,][]{smith2009b},
while \citet{fox2011} prefer a more massive progenitor like a LBV.
Our results seem to support the latter scenario but depend on
the value of $\epsilon$ assumed in deriving $D$ so we cannot constrain the
progenitor strongly (see Section \ref{sensitivity}).
In principle, we may be able to distinguish between the two progenitors
with $n$, but our results are found not to depend much on $n$.
Binary evolution may also be related to the dense CSM
\citep[e.g.,][]{chevalier2012}.

\subsubsection{SN 2006jd}
SN 2006jd was also observed in a wide spectral range by
\citet{stritzinger2012} and they obtained a bolometric LC.
We use their bolometric LC for our modeling.
We assume that the date of the explosion is 9 days before
its discovery and the following dates are since the explosion.
\citet{chandra2012} estimate CSM properties of SN 2006jd
based on the X-ray and radio observations
after about 400 days since explosion.
They conclude that
the CSM density profile is rather flat ($s\simeq 1.5-1.6$) and
the CSM density is $\sim 10^6~\mathrm{cm^{-3}}$
at $\sim 2\times 10^{16}$~cm.
\citet{fox2011,fox2013} estimate the mass-loss rate based on near-infrared
observations ($2.8\times 10^{-3}~M_\odot~\mathrm{yr^{-1}}$).

By fitting the LC before 250 days with $L=L_1 t^\alpha$, we obtain
\begin{equation}
L=3.87\times 10^{42}\left(\frac{t}{\mathrm{1~day}}\right)^{-0.0708}~\mathrm{erg~s^{-1}}.
\end{equation}
From $\alpha=-0.0708$ obtained by the bolometric LC fitting,
we obtain $s=1.4$ and $s=1.6$ for $n=10$ and $n=12$, respectively.
The shell velocity of SN 2006jd
around 100 days since the explosion is likely
$\simeq 15,000~\mathrm{km~s^{-1}}$ \citep{stritzinger2012}.
Then, based on Equation (\ref{generalD}), we get
\begin{equation}
\rho_\csm(r)=\left\{ \begin{array}{ll}
2.6\times 10^{-16}\left(\frac{r}{10^{15}~\mathrm{cm}}\right)^{-1.4}~\mathrm{g~cm^{-3}} & (n=10), \\ 
4.8\times 10^{-16}\left(\frac{r}{10^{15}~\mathrm{cm}}\right)^{-1.6}~\mathrm{g~cm^{-3}} & (n=12). \\ 
\end{array} \right.
\end{equation}
The Thomson scattering optical depth $\tau_\s$ of the solar-metallicity
unshocked CSM when the shell is at the radius $10^{15}$ cm is
\begin{equation}
\tau_\s=\left\{ \begin{array}{ll}
0.22 & (n=10), \\ 
0.26 & (n=12), \\ 
\end{array} \right.
\end{equation}
and our model is self-consistent.
We estimate an average mass-loss rate by using the CSM mass within $10^{16}$ cm
and $v_\w=100~\mathrm{km~s^{-1}}$ as we did for SN 2005ip in the
previous section.
The average mass-loss rate is
\begin{equation}
\langle \dot{M}\rangle=\left\{ \begin{array}{ll}
1.3\times 10^{-3}~M_\odot~\mathrm{yr^{-1}} & (n=10), \\ 
1.7\times 10^{-3}~M_\odot~\mathrm{yr^{-1}} & (n=12). \\ 
\end{array} \right.
\end{equation}
The estimated average mass-loss rate is consistent with the rate
derived by \citet{fox2011} from dust emission ($2.8\times 10^{-3}~M_\odot~\mathrm{yr^{-1}}$).
Interestingly, the mass-loss rate is very close to those of SN 2005ip
estimated in the previous section, although the density slopes are
quite different ($s=2.3-2.4$ for SN 2005ip and $s=1.4-1.6$ for SN 2006jd).

The late phase LC of SN 2006jd shows an increase
which is not expected in our model so we do not use the late time LC
in the fit.
This luminosity increase may be due to, e.g., another CSM component.
Since we can only fit the early phases,
we cannot constrain $M_\ej$ and $E_\ej$ independently.
Here, we assume $M_\ej=10~M_\odot$ to estimate $E_\ej$.
The estimated $E_\ej$ is
\begin{equation}
E_\ej=\left\{ \begin{array}{ll}
1.2\times 10^{52}~\mathrm{erg} & (n=10), \\ 
1.3\times 10^{52}~\mathrm{erg} & (n=12). \\ 
\end{array} \right.
\end{equation}
Note again that we assume $\epsilon=0.1$ (see Section \ref{sensitivity}).
The time $t_t$ obtained by these values are
\begin{equation}
t_t=\left\{ \begin{array}{ll}
4.1\times 10^2~\mathrm{days} & (n=10), \\ 
1.8\times 10^2~\mathrm{days} & (n=12). \\ 
\end{array} \right.
\end{equation}
The epochs we used to fit $L=L_1t^\alpha$ ($t<250$ days) 
are justified for the $n=10$ case.
For the $n=12$ case, $t_t$ is smaller than 250
days. However, there are only two observational data points beyond 180 days
and we find that
the results of fitting by using $t<180$ days are almost the same
as the results we obtained with $t<250$ days.

The CSM properties we derived are consistent with $s\simeq 1.5-1.6$ 
and the CSM density $\sim 10^6~\mathrm{cm^{-3}}$  at $\sim 2\times 10^{16}$ cm
as obtained by \citet{chandra2012}
from X-ray and radio observations.
However, the X-ray and radio observations were performed after
the epochs when the bolometric LC starts to rise (after about 400 days
since the explosion).
Our model is not applicable at these epochs as is discussed above
and this correspondence can be a coincidence.

\begin{figure}
\begin{center}
 \includegraphics[width=0.49\columnwidth]{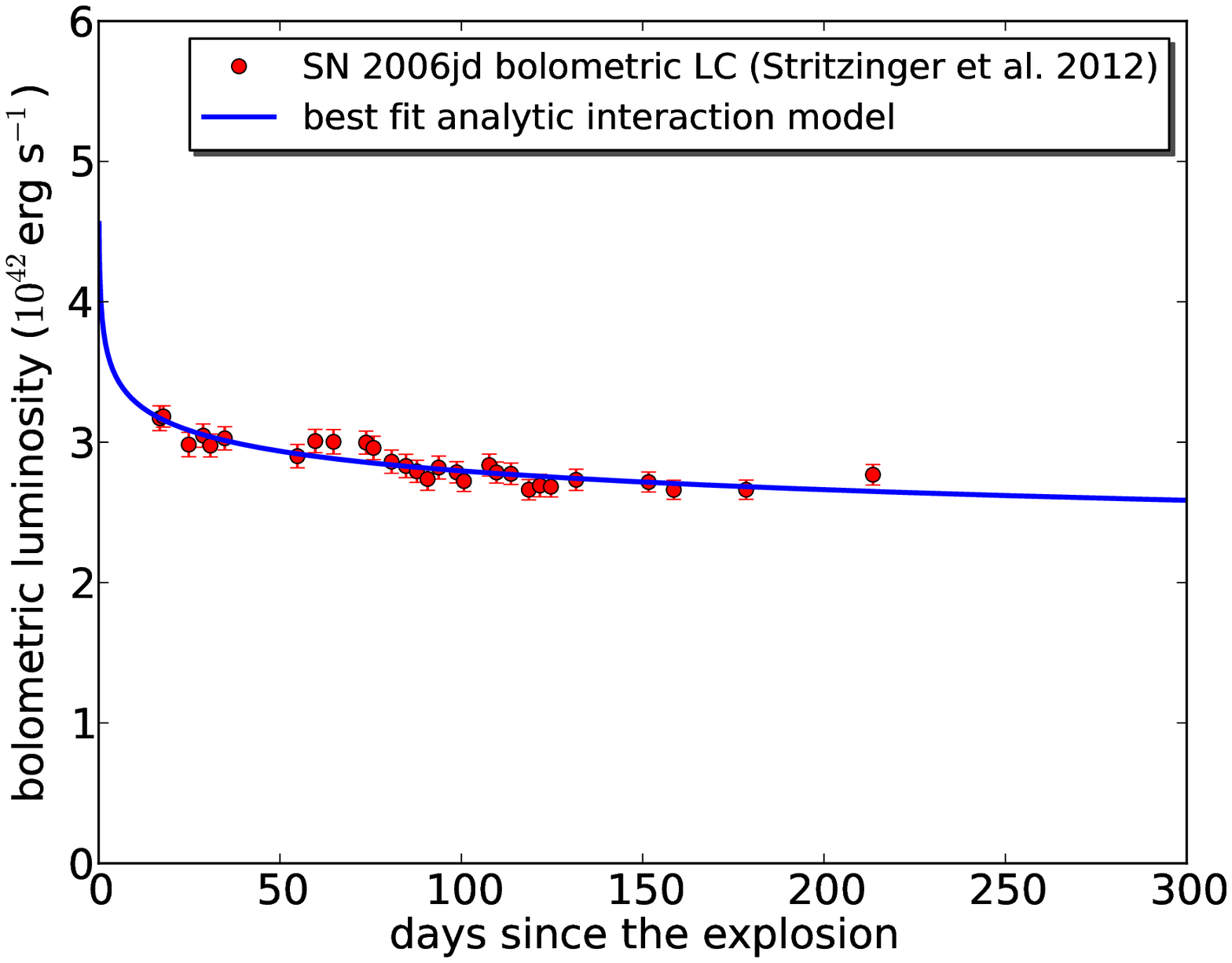}
 \includegraphics[width=0.49\columnwidth]{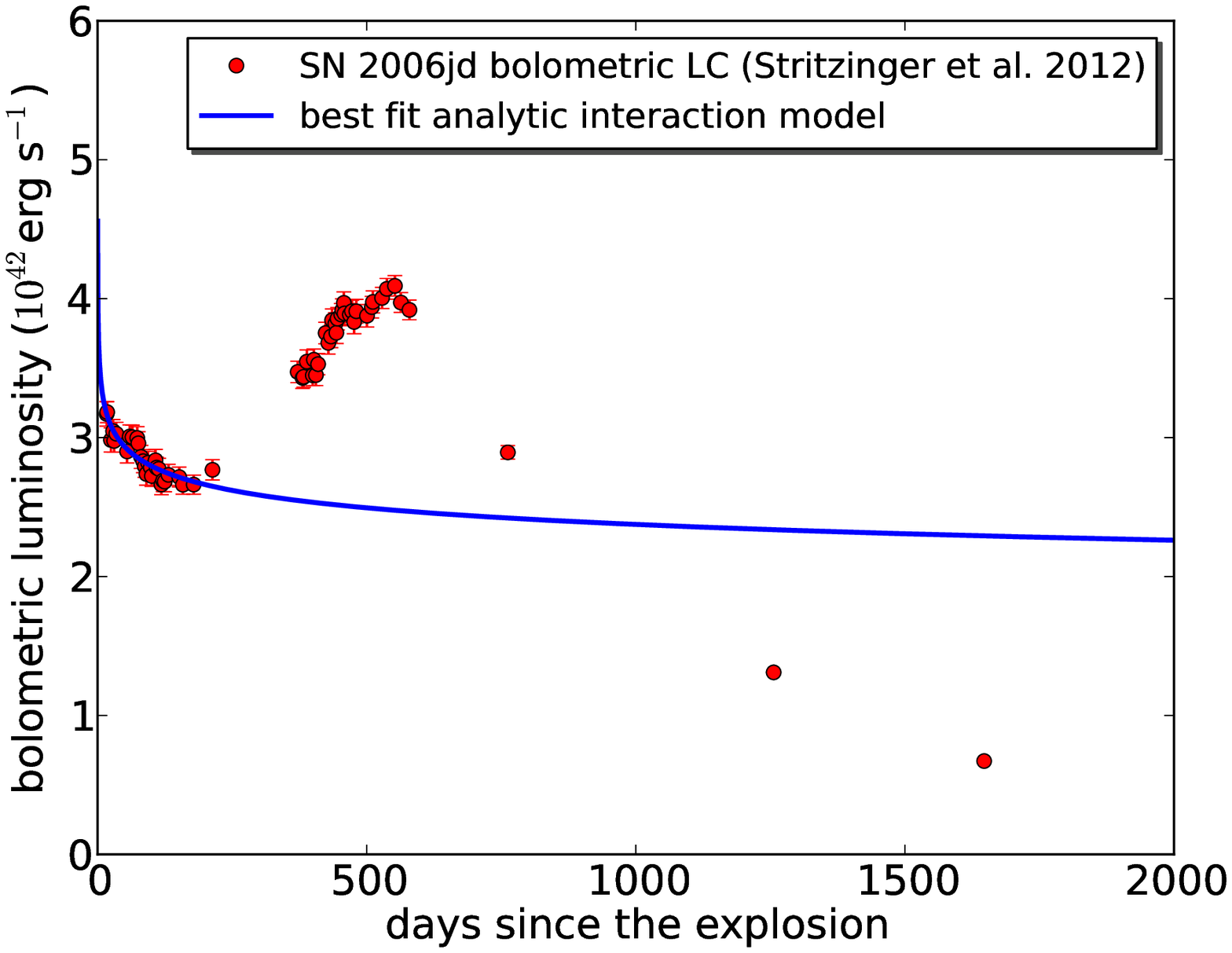}
  \caption{
Bolometric LC of SN 2006jd \citep{stritzinger2012} and the best fit
 $L=L_1 t^\alpha$ model up to 250 days.
The second rise starting around 400 days
cannot be explained by our model and may be due to, e.g.,
another CSM component.
}
\label{sn2006jd}
\end{center}
\end{figure}

\subsubsection{SN 2010jl}
SN 2010jl has been extensively observed in a wide range of wavelengths
\citep{smith2011,smith2012,stoll2011,andrews2011,chandra2012b,fox2013,maeda2013b}.
\citet{zhang2012} obtained a bolometric LC of SN 2010jl based on their
optical observations and we use it for our model comparison.
Note that they do not have near-infrared observations and the bolometric
LC is constructed without them.
The date of the explosion is set to 12 days before the $V$-band LC
peak reported by \citet{stoll2011}.
We apply our spherically symmetric bolometric LC model but the observation of
polarization indicates an asymmetric nature of the CSM around
SN 2010jl \citep{patat2011}.

At first, we use $L=L_1t^\alpha$ to fit the bolometric LC and get
\begin{equation}
L=2.04\times 10^{44}\left(\frac{t}{1~\mathrm{day}}\right)^{-0.486}~\mathrm{erg~s^{-1}}
\end{equation}
$\alpha=-0.486$ corresponds to $s=2.2$ and $s=2.3$ for $n=10$ and
$n=12$, respectively.
However, $t_t$ become
\begin{equation}
t_t=\left\{ \begin{array}{ll}
9.5~\mathrm{days} & (n=10), \\ 
3.8~\mathrm{days} & (n=12), \\ 
\end{array} \right.
\end{equation}
for $E_\ej=10^{52}~\mathrm{erg}$ or
\begin{equation}
t_t=\left\{ \begin{array}{ll}
57~\mathrm{days} & (n=10), \\ 
23~\mathrm{days} & (n=12), \\ 
\end{array} \right.
\end{equation}
for $E_\ej=2.5\times 10^{52}~\mathrm{erg}$
with the obtained $L_1$.
This means that the $L=L_1t^\alpha$ formula we used for the fitting
is not applicable for most of the epochs we used for the fitting.
Thus, we need to use the asymptotic formula (Equation \ref{asymgen}) to
fit the LC.

\begin{figure}
\begin{center}
 \includegraphics[width=0.75\columnwidth]{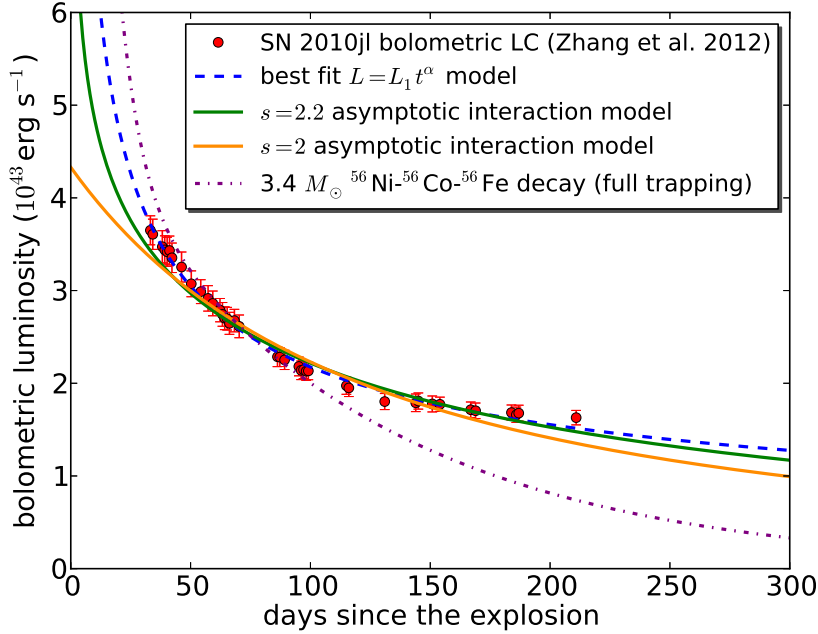}
  \caption{
Bolometric LC of SN 2010jl \citep{zhang2012} and some model fits to it.
The dashed line represents the best fit for $L=L_1t^\alpha$.
However, $t_t$ expected from the result of the fit is too small
to apply this model to the entire LC. Thus, we need to apply
the asymptotic LC formula which is applicable after $t_t$.
We get a good fit with $s=2.2$ (green line).
The orange line is the best fit from the $s=2$ model.
The dot-dashed line is the radioactive decay energy available
from $3.4~M_\odot$ \Ni. The radioactive decay model is suggested
by \citet{zhang2012} to explain the early bolometric LC but
our interaction model can explain the entire LC with a single component.
}
\label{sn2010jl}
\end{center}
\end{figure}

In Figure \ref{sn2010jl}, we show some asymptotic LC models from
Equation (\ref{asymgen}). 
We have searched for a good fit by changing $s$,
$D$, and $E_\ej$. We assume $M_\ej=10~M_\odot$.
The best model we found is shown
in Figure \ref{sn2010jl} and it has
\begin{equation}
\rho_\csm(r)=2.5\times
 10^{-14}\left(\frac{r}{10^{15}~\mathrm{cm}}\right)^{-2.2}~\mathrm{g~cm^{-3}},
\end{equation}
 and
\begin{equation}
E_\ej=2.3\times 10^{52}\ \mathrm{erg}.
\end{equation}
The Thomson scattering optical depth of the solar-metallicity unshocked
CSM when the shell is at $10^{15}$ cm is $\tau_\s=7.1$.
$\tau_\s$ becomes $\sim 1$ at $\sim 5\times 10^{15}$ cm and the shell is
above $\sim 5\times 10^{15}$ cm at the epochs we apply our model
(after about 30 days since the explosion).
The average mass-loss rate estimated by the CSM mass within $10^{16}$ cm
for $v_\w=100~\mathrm{km~s^{-1}}$ is
\begin{equation}
\langle\dot{M}\rangle = 0.039~M_\odot~\mathrm{yr^{-1}}.
\end{equation}
The estimated rate is consistent with those obtained based on the
infrared emission \citep{maeda2013b,fox2013}.
$t_t=29$ days for $n=10$ and $\delta=1$ and $t_t=15$ days
for $n=12$ and $\delta=1$.
Thus the usage of the asymptotic formula is justified.

Since $s$ obtained above is close to the case of the steady mass loss ($s=2$),
we also try to fit the bolometric LC by the asymptotic formula
$L=a(1+bt)^{-3/2}$ for $s=2$ (Equation \ref{latee}).
We obtain
$a=4.38\times 10^{43}~\mathrm{erg~s^{-1}}$ and
$b=6.44\times 10^{-8}~\mathrm{s^{-1}}$
with $t_t=22$ days ($n=10$ and $\delta=1$) or $t_t=13$ days ($n=12$ and $\delta=1$).
By using $a$, $b$, and $\epsilon=0.1$, we get
\begin{eqnarray}
E_\ej=1.4\times10^{52}~\mathrm{erg},
\end{eqnarray}
from Equation (\ref{Eej}).
Assuming $M_\ej=10~M_\odot$ and $v_w=100~\mathrm{km~s^{-1}}$,
we obtain
\begin{equation}
\dot{M}=0.087~M_\odot~\mathrm{yr^{-1}}, 
\end{equation}
from Equation (\ref{Mej}). The rate is similar to the average rate
from the $s=2.2$ model derived above.

Comparing the $s=2.2$ and $s=2$ models, we find that
the $s=2$ model has flatter LC than the $s=2.2$ model.
As we make $s$ smaller, the model LC gets flatter and
it gets harder to explain the bolometric LC of SN 2010jl.
Thus, we presume that the CSM around the progenitor of SN 2010jl
may be a bit steeper than the CSM expected from steady mass loss.
This conclusion contradicts that obtained by \citet{chandra2012b} from X-ray observations.
\citet{chandra2012b} suggest $s=1.6$ for SN 2010jl based on X-ray observations.
However, their estimate is obtained by assuming $r_\s\propto t^{(n-3)/(n-s)}$
which is not likely applicable at the epochs when they obtained X-ray
data ($\simeq 60$ days and $\simeq 360$ days since the explosion).
This is because of the small $t_t$ mainly due to the high CSM density as is shown above.

So far, we fit the entire bolometric LC up to about 200 days by a
single component. On the contrary, \citet{zhang2012} suggested
a two-componet model for the bolometric LC.
They suggested that the LC before around 100 days is mainly powered by
$3.4~M_\odot$ of \Ni\ whose available radioactive energy is shown in Figure
\ref{sn2010jl}. They suggested that the SN ejecta-CSM interaction
started playing a role at later epochs by using a model LC of the interaction
developed by \citet{wood-vasey2004}.
However, the required \Ni\ mass is very large $(3.4~M_\odot)$
and this amount of \Ni\ is rather difficult to produce in
a core-collapse SN explosion \citep[e.g.,][]{umeda2008}.
In addition, no signatures of Fe elements are observed in the late phase
spectra of SN 2010jl which are expected if there is large amount of
\Ni\ production \citep[e.g.,][]{dessart2013}. 
As noted in Section \ref{sec:05ip}, the bolometric LC powered by
the interaction resembles to the LC powered by the radioactive decay
of \Ni\ at early epochs
and we need to use additional late-phase LCs to distinguish between the two power sources.
We have shown here that we need only one component from the interaction model
to explain the whole LC of SN 2010jl.

\section{Discussion}\label{sec4}

\subsection{Sensitivity to the Assumed Parameters}\label{sensitivity}
We have fixed $\epsilon=0.1$ and $M_\ej=10~M_\odot$
in deriving some SN ejecta and CSM properties in the previous section.
Here, we discuss how sensitive the derived properties are to
the assumed values of $\epsilon$ and $M_\ej$.
In addition, it is practically difficult determine $v_1$ observationally from
spectra of SNe IIn because the origins of the spectral features are
not understood well. We also discuss the effect of the uncertainty in $v_1$.

At first, we assume that $v_1$ is well-determined and see the effect of
$\epsilon$. The estimated CSM density structures or the estimated
mass-loss rates depend on $\epsilon$ for a given $v_1$
(Equation \ref{generalD}). The average mass-loss rates have a relation
$\langle\dot{M}\rangle\propto \epsilon^{-1}$.
For example, we obtained
$\langle\dot{M}\rangle=1.2\times 10^{-3}\ M_\odot~\mathrm{yr^{-1}}$
for SN 2005ip ($n=10$) in the previous section by assuming $\epsilon=0.1$
but this mass-loss rate may be reduced to 
$\langle\dot{M}\rangle=2.4\times 10^{-4}\ M_\odot~\mathrm{yr^{-1}}$
if we assume $\epsilon=0.5$.
This uncertainty makes it difficult to distinguish the RSG progenitor
and the LBV progenitor.
The assumption of $M_\ej=10~M_\odot$ is used to estimate $E_\ej$.
$E_\ej$ depends on both $M_\ej$ and $\epsilon$ as
$E_\ej\propto \epsilon^{-\frac{2}{n-3}}M_\ej^{\frac{n-5}{n-3}}$.
For the typical values of $n$, we obtain
$E_\ej\propto \epsilon^{-0.29}M_\ej^{0.71}$ $(n=10)$
or
$E_\ej\propto \epsilon^{-0.22}M_\ej^{0.78}$ $(n=12)$.
In either case, $E_\ej$ is mostly determined by the assumed $M_\ej$.
For the SN 2005ip model ($n=10$) in the previous section
($E_\ej=1.3\times 10^{52}$ erg for $\epsilon=0.1$ and $M_\ej=10~M_\odot$),
$E_\ej$ can be changed to, e.g.,
$7.9\times 10^{51}$ erg ($\epsilon=0.1$ and $M_\ej=5~M_\odot$),
$8.2\times 10^{51}$ erg ($\epsilon=0.5$ and $M_\ej=10~M_\odot$), or
$5.0\times 10^{51}$ erg ($\epsilon=0.5$ and $M_\ej=5~M_\odot$).
Thus, the estimated $E_\ej$ are not much affected by the assumed parameters.

We have assumed that $v_1$ can be determined by spectral
observations. However, spectra of SNe IIn have complicated
features with several components and it is not obvious which spectral
component originates from the dense shell and can be used to estimate $v_1$.
Thanks to the formation of the dense shell due to the radiative cooling,
the shell velocity is one of
the fastest components in the system (see also Section \ref{secnumerical}).
Thus, we have used the fastest velocity component in the spectra to
estimate $v_1$ in the previous section
($17,500~\mathrm{km~s^{-1}}$ at 100 days for SN 2005ip).
Indeed, $\langle\dot{M}\rangle\propto v_1^{s-5}$ 
($\langle\dot{M}\rangle\propto v_1^{-3}$ for $s\simeq 2$)
and the estimated mass-loss rates are rather sensitive to the assumed $v_1$.
Keeping $\epsilon=0.1$ and $M_\ej=10~M_\odot$, we obtain 
$\langle\dot{M}\rangle=1.8\times 10^{-3} M_\odot~\mathrm{yr^{-1}}$
and $E_\ej=1.0\times 10^{52}$ erg
for $v_\s(\mathrm{100~days})=15,000\ \mathrm{km~s^{-1}}$,
$\langle\dot{M}\rangle=5.5\times 10^{-3} M_\odot~\mathrm{yr^{-1}}$
and $E_\ej=5.8\times 10^{51}$ erg
for $v_\s(\mathrm{100~days})=10,000\ \mathrm{km~s^{-1}}$, and
$\langle\dot{M}\rangle=3.6\times 10^{-2} M_\odot~\mathrm{yr^{-1}}$
and $E_\ej=2.2\times 10^{51}$ erg
for $v_\s(\mathrm{100~days})=5,000\ \mathrm{km~s^{-1}}$.
As we use the highest velocity component to estimate $v_1$ in the
previous section, the estimated mass-loss rates were rather conservative.
Note again that the dense shell velocity should be one of the fastest components
in the system we model (see also Section \ref{secnumerical}) and
thus adopting the higher observed velocity is preferred.

Finally, $v_\w$ has been assumed to be $100~\mathrm{km~s^{-1}}$, which
is a typical LBV wind velocity \citep[e.g.,][]{leitherer1997}.
SN IIn spectra often show a $100~\mathrm{km~s^{-1}}$ P-Cygni profile.
The mass-loss rates estimated are proportional to $v_\w$.
If SNe IIn are from RSGs, the wind velocity can be lower
\citep[e.g.,][]{mauron2011} and the
mass-loss rates estimated will be decreased because of the lower wind velocities.

\subsection{Applicability}
In deriving the evolution of the shocked-shell radius $r_\s(t)$,
we have assumed that $s$ is smaller than 3.
This condition is also required to derive a physical self-similar solution
\citep[e.g.,][]{nadyozhin1985}.
The allowed range of $\alpha$ for $s<3$ is $\alpha>-1$
because $\alpha\rightarrow -1$ $(s\rightarrow 3)$ and 
$\alpha$ is a monotonically-decreasing function at $n>5$.
Thus, if we obtain $\alpha<-1$ by fitting $L=L_1t^\alpha$,
this is beyond the applicability of our model and we need
to consider other ways to explain the LC.

First, we need to check $t_t$. If $t_t$ is smaller than the time
used for the fitting, we need to use the asymptotic formula
for the fitting. The asymptotic formula can have a rapid decline
in the bolometric LC depending on parameters.

Another possibility is a CSM with $s>3$. 
Most of the mass in CSM with $s>3$ exists near the inner edge of
the CSM. In other words, for the case of $s>3$,
\begin{eqnarray}
M_\csm&\equiv&\int^{r_\s}_{R_p}4\pi r^2 \rho_\csm dr \\
&=&\frac{4\pi D}{s-3}\left(R_p^{3-s}-r_\s^{3-s}\right) \\
&\simeq&\frac{4\pi D}{s-3}R_p^{3-s}=\mathrm{constant}\ \ (r_\s\gg R_p).
\end{eqnarray}
Thus, most of the CSM is shocked soon after the explosion.
If the CSM density is relatively low, the LCs will decline quickly
soon after the explosion when most of the CSM component is swept up.
If the shocked shell becomes optically thick,
LCs may resemble the 'shell-shocked diffusion' model LC
suggested by \citet{smith2007} as a model for superluminous SNe
based on the formalisms by \citet{arnett1980} \citep[but see also][]{moriya2013}.
This is a LC model for the declining part of the bolometric LC
after the shock wave passes through a dense CSM.
According to this model, the declining part of the bolometric LCs follows
\begin{equation}
L=L_0\exp\left[-\frac{t}{\tau_\mathrm{diff}}\left(1+\frac{t}{2\tau_\mathrm{exp}}\right)\right],
\label{shockedshell}
\end{equation}
where $t$ is the time since the maximum luminosity, $\tau_\mathrm{diff}$
is the characteristic diffusion timescale in the shocked shell and
$\tau_\mathrm{exp}$ is the expansion timescale of the shocked shell.

Bolometric LCs can also follow Equation (\ref{shockedshell})
even if $s<3$.
This is the case when the high-density CSM is 
small in radius and the entire high-density CSM is shocked
soon after the beginning of the interaction.
Then, there is no continuous interaction and the bolometric LC
should decline quickly, possibly following the shell-shocked diffusion model.
However, in this case, there may be little remaining CSM to emit narrow
emission lines in spectra and the SN may not continue to be of Type IIn.

So far, we have only considered possible ways to understand rapidly declining LCs in the
context of the SN ejecta-dense CSM interaction.
For the case of SNe IIn, it is natural to consider in the context of the interaction model.
However, it is possible that CSM around some SNe IIn are
dense enough only to affect their spectra while their LCs are not much affected
by the dense CSM. Then, rapidly declining LCs may be powered by other mechanisms
like \Ni, magnetars \citep[e.g.,][]{maeda2007,kasen2010b,woosley2010}, or fallback \citep{dexter2012}.

\subsection{Initial Luminosity Increase}
The bolometric LC model presented in this paper
does not have a rising part at the beginning.
There are several mechanisms to make the initial luminosity increase
in LCs which are not taken into account in our model.

We have assumed that the radiation emitted from the dense shell
is not affected by the unshocked CSM.
However, especially at the early phases just after the explosion,
the CSM surrounding the dense shell can be optically thick
and the radiation from the shell can be scattered within the CSM.
In this case,
the diffusion timescale in the optically thick region determines
the evolution of the initial luminosity increase and
subsequent decline.
Our model should only be applied to the epochs when the CSM surrounding
the dense shell becomes optically thin and should not be applied
at the epochs when the luminosity increases or just after the luminosity peaks.
When the CSM is optically thick, some signatures can be
seen in spectra as well \citep[e.g.,][]{chugai2001}.

If the CSM is optically thin, the timescale of the initial luminosity
increase is expected to be very small. 
Two mechanisms can affect the initial luminosity increase.
One is the shock breakout at the surface of the progenitor and
the other is the on-set of the SN ejecta-CSM interaction.
Both are presumed to have a short timescale.
If the dense part of the CSM and the progenitor are detached,
we may see two luminosity peaks in the early phases:
one from the shock breakout and the other from the on-set of the interaction.

\begin{figure}
\begin{center}
 \includegraphics[width=0.6\columnwidth]{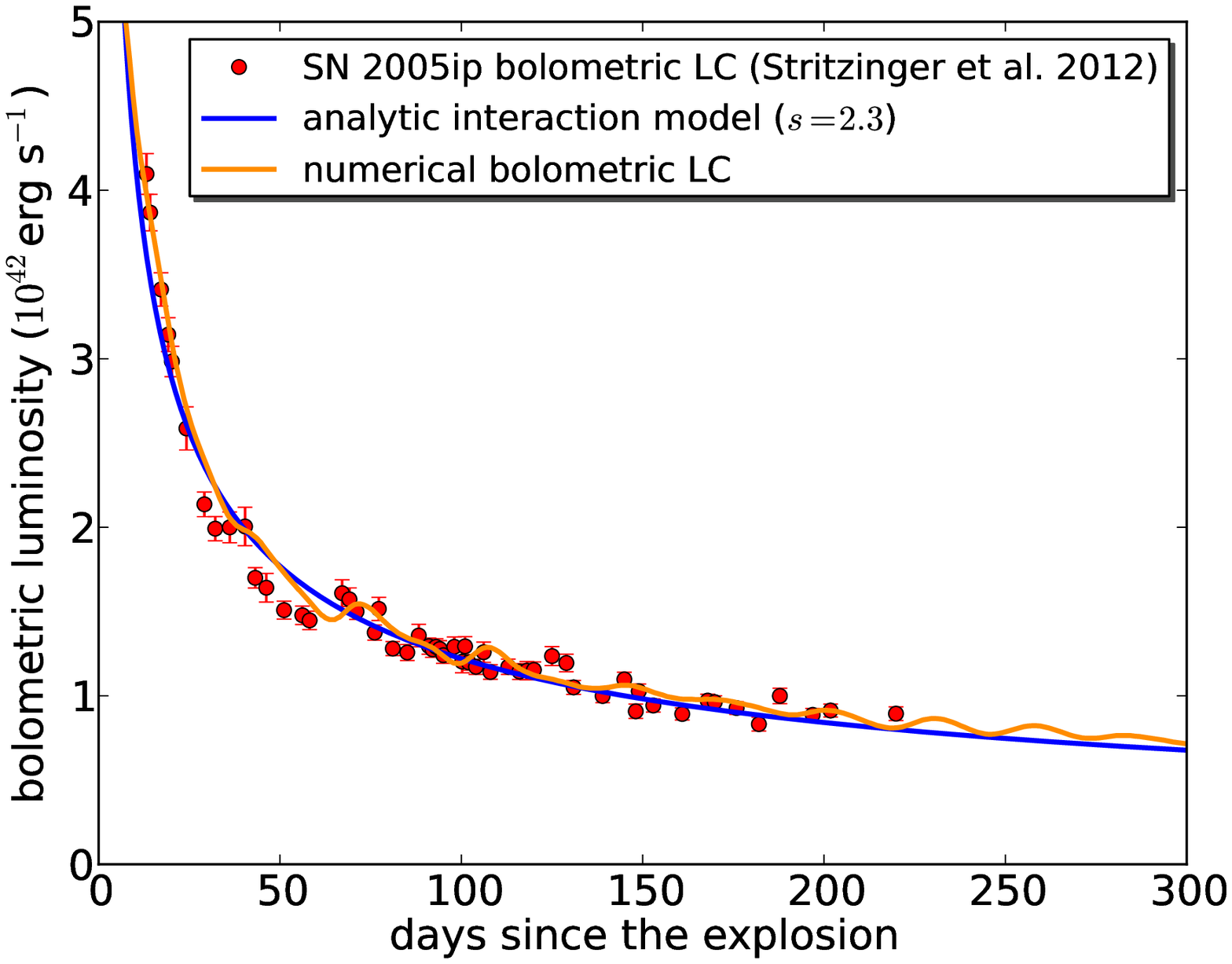} \\
 \includegraphics[width=0.6\columnwidth]{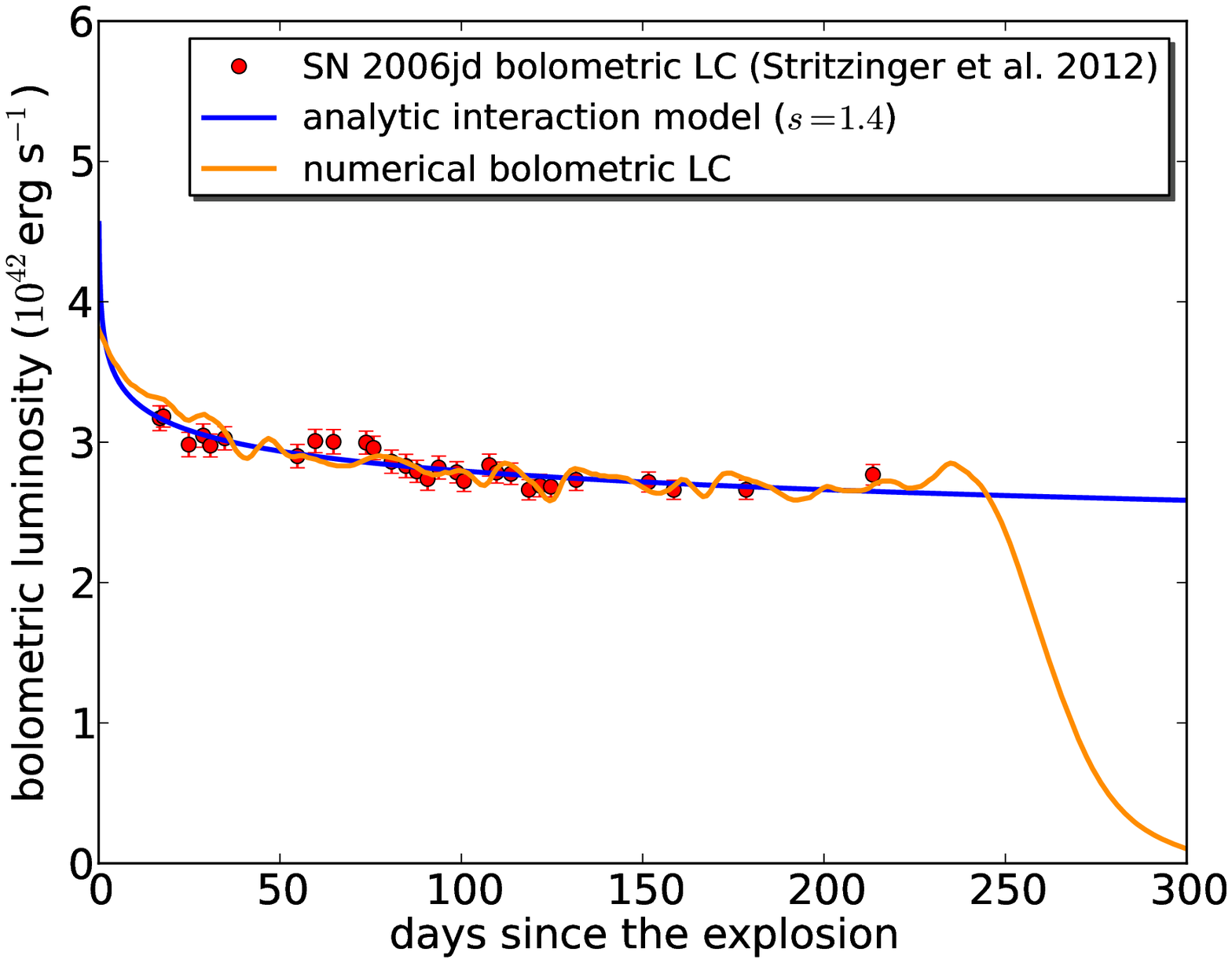} \\
 \includegraphics[width=0.6\columnwidth]{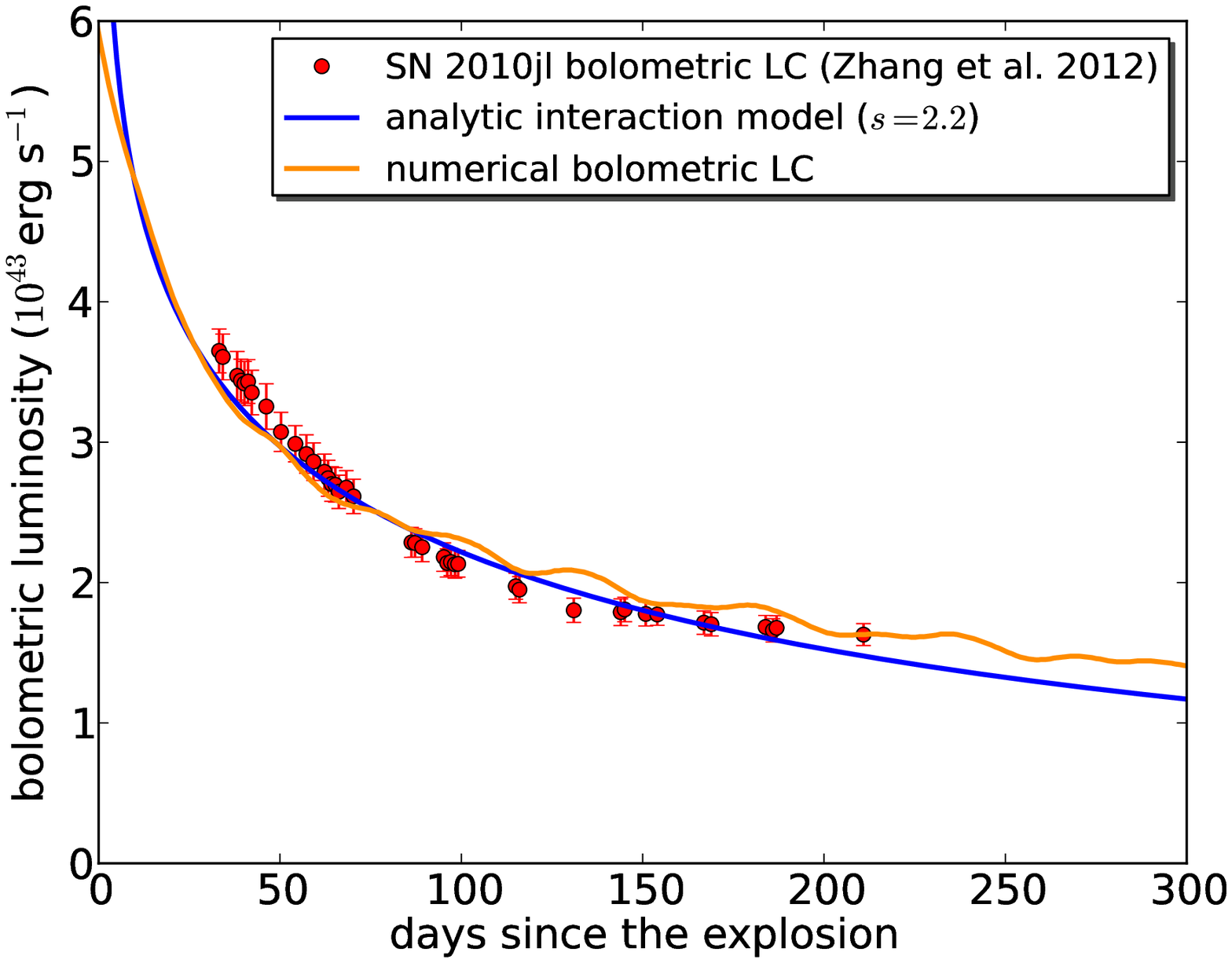}
 \caption{
Comparisons between the numerical bolometric LCs and the analytic
bolometric LCs presented in Section \ref{exmpls}
based on which the initial conditions
for the numerical bolometric LC computations are constructed.
}
\label{numerical}
\end{center}
\end{figure}

\begin{figure}
\begin{center}
 \includegraphics[width=0.49\columnwidth]{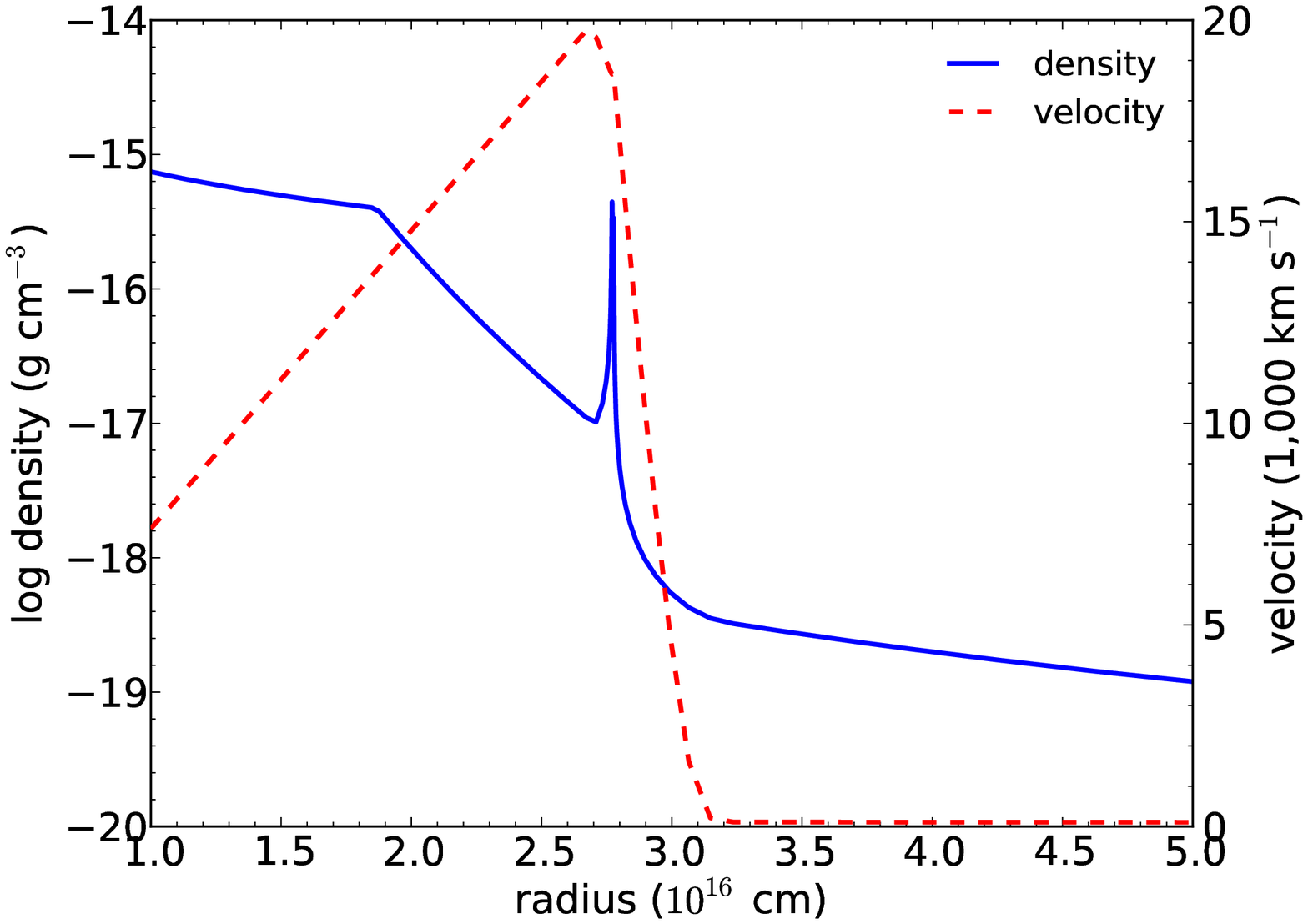} 
 \includegraphics[width=0.49\columnwidth]{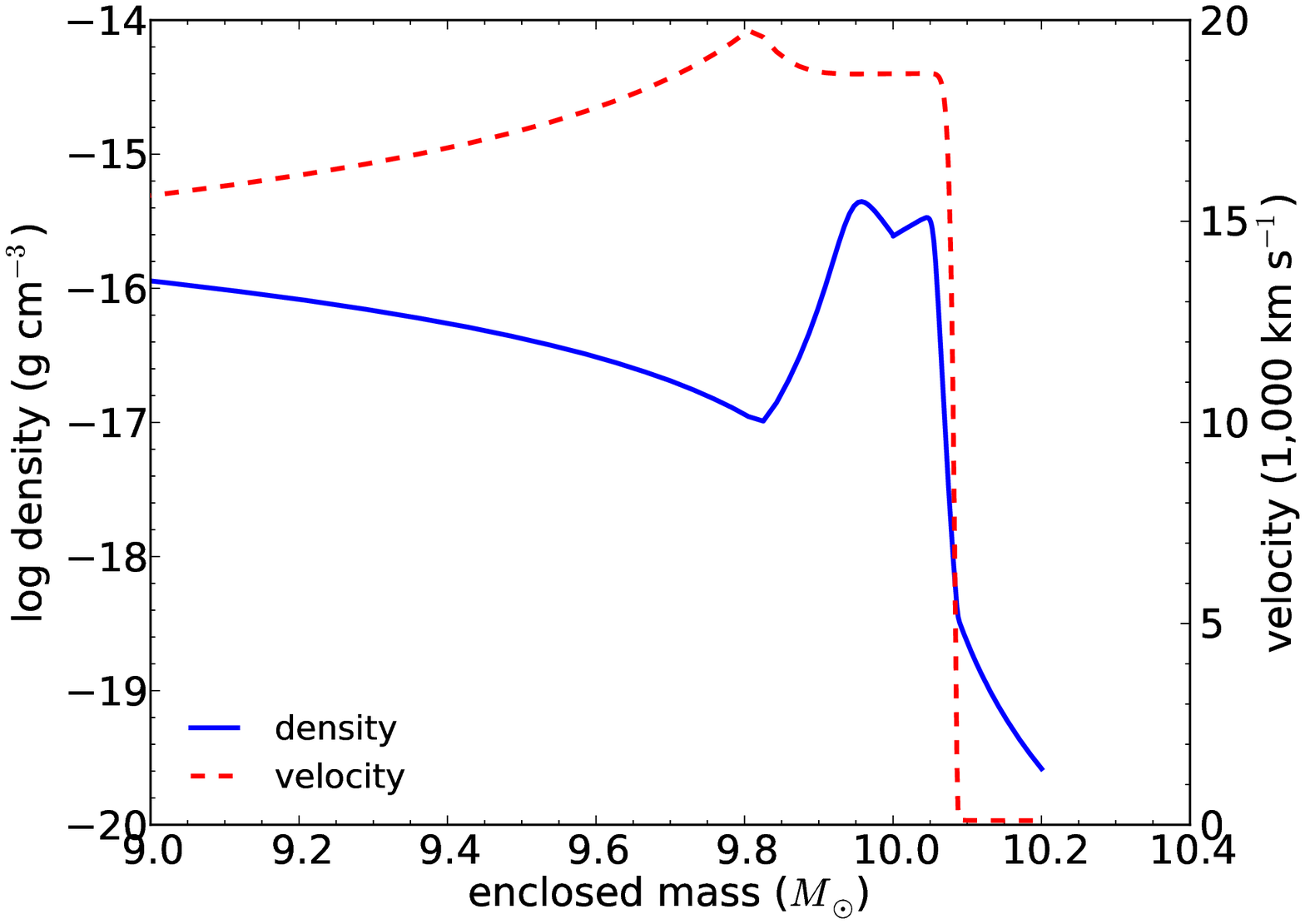}
 \caption{
Density and velocity structures of the numerical model for SN 2005ip
at 100 days.
}
\label{numericalhyd}
\end{center}
\end{figure}

\subsection{Comparison with Numerical Bolometric Light Curves}\label{secnumerical}
To show the reliability of our analytic LC model,
we also performed numerical LC calculations using 
a one-dimensional radiation hydrodynamics code \verb|STELLA|
\citep[e.g.,][]{blinnikov1993,blinnikov2000,blinnikov2006}.
We show some comparisons in this paper
and more detailed comparisons will be presented elsewhere.
We set the initial conditions following the physical parameters
obtained in Section \ref{exmpls}. The density structure of the
 homologously expanding SN ejecta has two power-law components
as is assumed in the analytic model. The SN ejecta and CSM
are initially connected at $10^{14}$ cm.
The CSM outer radius of all the models is set to $10^{17}$ cm.
The parameter $B_q$ which controls the conversion efficiency 
from the kinetic energy to radiation in the code
(see \citealt{moriya2013b} for the details)
is adjusted to make $\epsilon\simeq0.1$.
Both SN ejecta and CSM in the calculations have solar composition.

Figure \ref{numerical} presents the results of our LC calculations.
We performed the LC calculations of three models in Section \ref{exmpls},
namely,
SN 2005ip ($s=2.3$, $n=10$, $\delta=1$, $E_\ej=1.2\times 10^{52}$ erg,
and $M_\ej=10~M_\odot$),
SN 2006jd ($s=1.4$, $n=10$, $\delta=1$, $E_\ej=1.3\times 10^{52}$ erg,
and $M_\ej=10~M_\odot$), and
SN 2010jl ($s=2.2$, $n=10$, $\delta=1$, $E_\ej=2.3\times 10^{52}$ erg,
and $M_\ej=10~M_\odot$).
The overall features of the analytic LCs
are well reproduced by the numerical LCs and the
analytic model presented in this paper is shown to provide a good
prediction to the numerical results.

In Figure \ref{numericalhyd}, we show the density and velocity
structures of the numerical SN 2005ip model at 100 days
in radius and mass coordinates.
We can see that the dense shell is formed between the SN ejecta and the dense
CSM and the shell width is much smaller than the shell radius
because of the radiative cooling.
The plot in the mass coordinate indicates that most of the shocked SN
ejecta and CSM is in this thin shell.
Thus the assumption in our analytic model that the shocked region
can be expressed by using a single $r_\s$ is verified.
This means that the forward and reverse shocks are glued to the cool
dense shell and the velocities of them
are not different from each other so much at these early epochs
because of the radiative cooling.
Note that the shell is one of the fastest velocity components
at this epoch and the shell velocity is consistent with $v_1$ adopted
($v_\s\simeq17,500\ \mathrm{km~s^{-1}}$ at 100 days).
We can also see that the density structure
ahead of the shell is modified slightly because of the precursor.

\subsection{Non-Bolometric Light Curves}
As our LC model takes only the sum of the available energy into account,
the LC we obtain from the model is bolometric and we have applied
our analytic bolometric LC model to bolometric LCs constructed from
observations.
Here we try to fit the $L=L_1t^\alpha$ formula to optical and near-infrared LCs
of SN 2005ip and SN 2006jd obtained by \citet{stritzinger2012}.
We focus on the parameter $\alpha$ which is directly affected by
the CSM density slope $s$ for a given $n$.

\begin{figure}
\begin{center}
 \includegraphics[width=0.49\columnwidth]{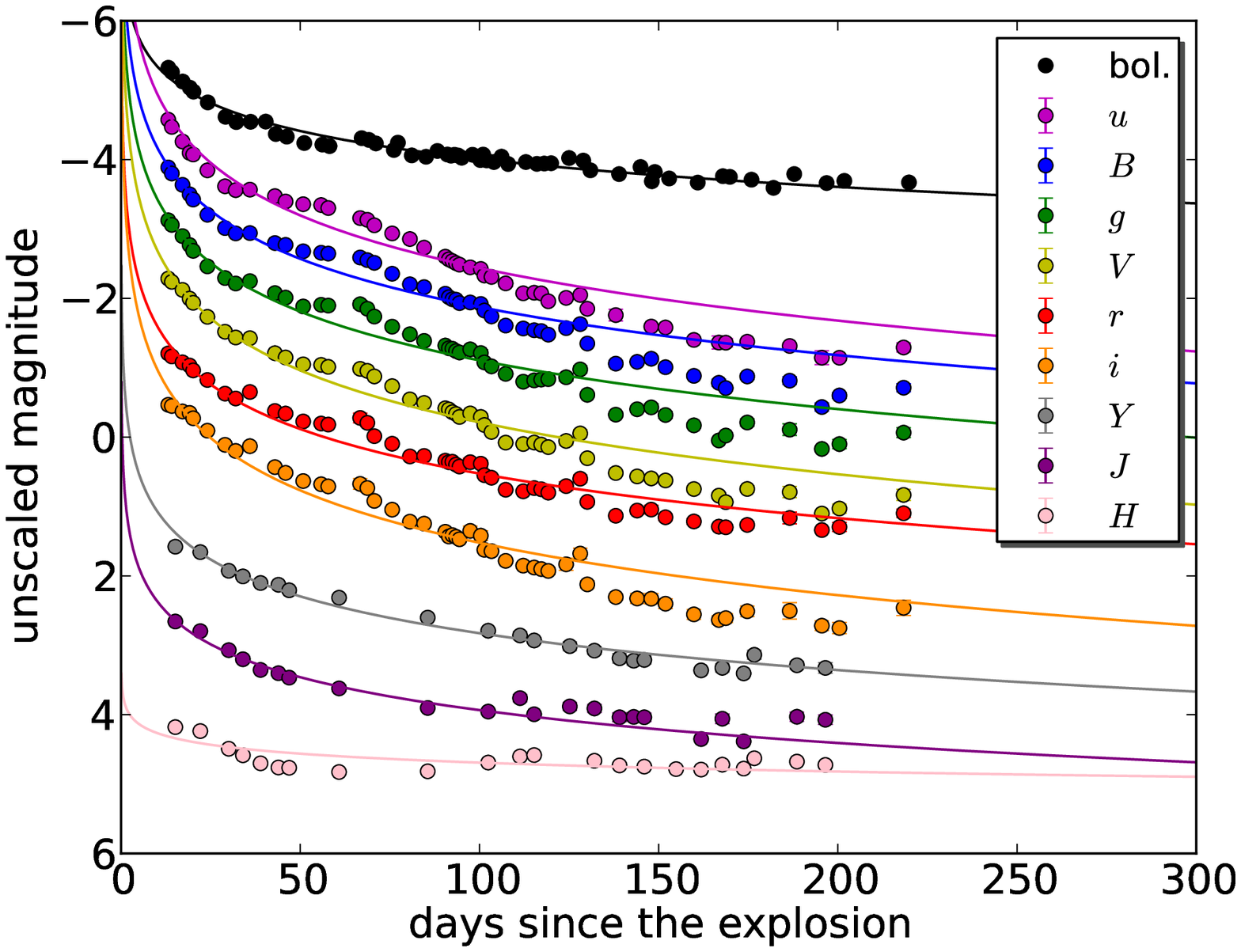}
 \includegraphics[width=0.49\columnwidth]{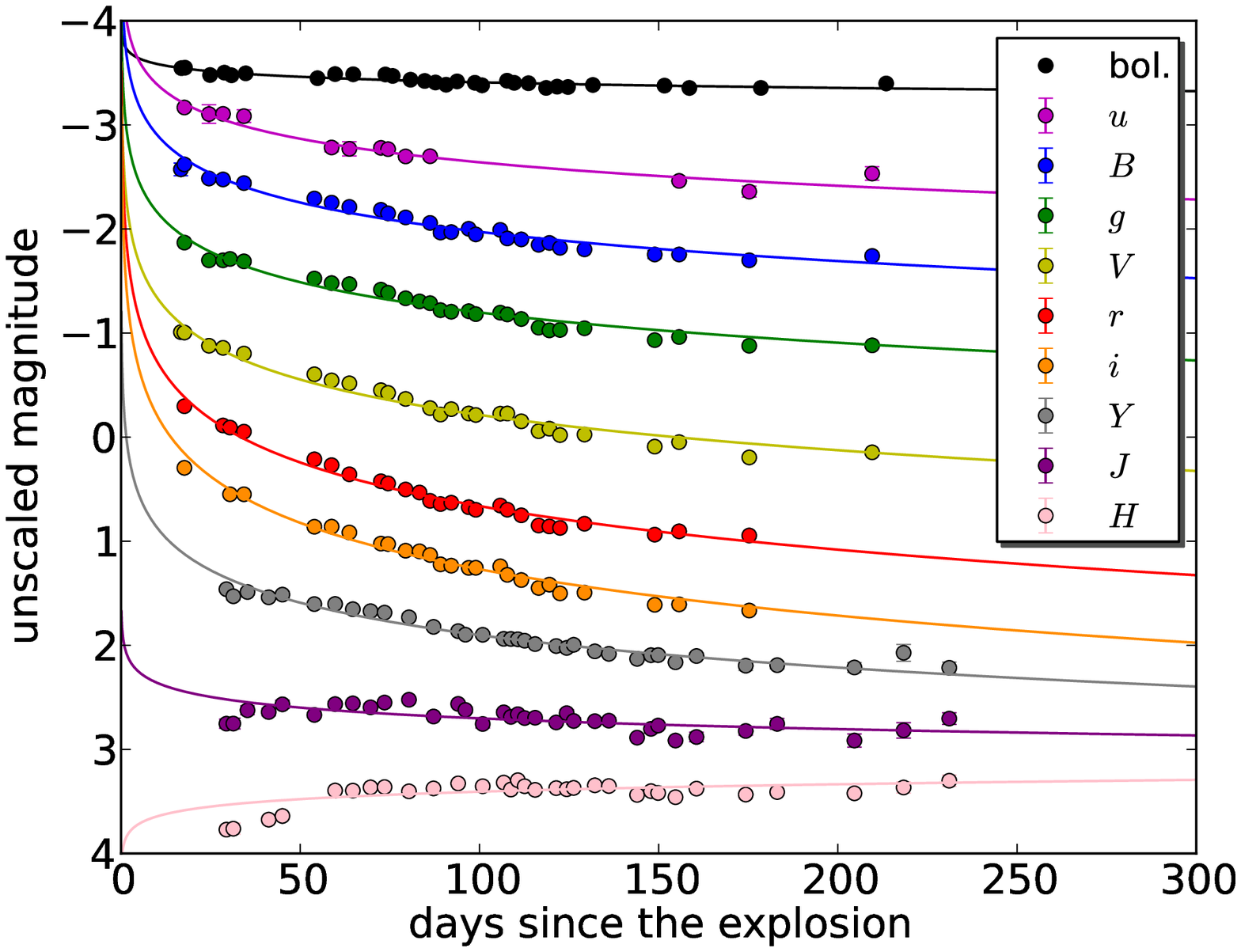}
 \caption{
Multi-color LCs of SN 2005ip and SN 2006jd and the results of the
 fit to $L\propto t^\alpha$.
$\alpha$ obtained by multi-color LCs are not consistent with $\alpha$
obtained from the bolometric LC. We need a bolometric LC to infer
CSM and SN ejecta properties from LCs properly.
}
\label{colorLC}
\end{center}
\end{figure}

Figure \ref{colorLC} and Table \ref{table1} show the results of the LC fits.
As we can see, $\alpha$ obtained with different photometric 
bands have different values. This means that we need to construct
a bolometric LC from observations to obtain accurate information.
This can be clearly seen in Figure 7 of \citet{stritzinger2012}.
The spectra evolve significantly with time and
no single band can represent the entire evolution of the
bolometric LC. We thus clearly need to construct a bolometric LC to apply
our model to obtain CSM and SN properties of SNe IIn.

\section{Conclusions}\label{sec5}
We have developed an analytic bolometric LC model for SNe powered by the
interaction between SN ejecta and dense CSM. This model is suitable for
modeling SNe IIn.
We have analytically derived the evolution of the shocked dense shell
created by the interaction.
We obtain the bolometric LC evolution from the derived dense shell evolution.
Our model is not restricted to the CSM from steady mass loss.

We have applied our bolometric LC model to three SNe IIn whose
bolometric LCs have been constructed from observations, i.e.,
SN 2005ip, SN 2006jd, and SN 2010jl.
The results show that their CSM density slopes are close to
what is expected from the steady mass loss ($s=2$ where
$\rho_\csm\propto r^{-s}$)
but slightly deviate from it
($s\simeq 2.3-2.4$ for SN 2005ip, $s\simeq 1.4-1.6$ for SN 2006jd,
and $s\simeq 2.2$ for SN 2010jl).
The derived mass-loss rates are consistent with
LBVs (SN 2005ip: $\langle\dot{M}\rangle=0.0012-0.0014~M_\odot~\mathrm{yr^{-1}}$,
SN 2006jd: $\langle\dot{M}\rangle=0.0013-0.0017~M_\odot~\mathrm{yr^{-1}}$ and
SN 2010jl: $\langle\dot{M}\rangle=0.039~M_\odot~\mathrm{yr^{-1}}$).
We could not constrain SN ejecta properties strongly but
$E_\ej$ of all three SNe
likely exceeded $10^{52}$ erg if we assume that $M_\ej=10~M_\odot$
and that the conversion efficiency
from kinetic energy to radiation is 10\% $(\epsilon=0.1)$.

We have also found that the energy inputs from the interaction and
the radioactive decay of \Ni\ can be similar to each other up to about 100 days
since the explosion. We need to have LCs also at later phases
to distinguish between the two luminosity sources from LCs alone.

Our bolometric LC model can only be applied for $s<3$.
For $s>3$, we suggest that the shell-shocked diffusion model
proposed by \citet{smith2007} \citep[see also][]{moriya2013}
may be applied for some cases.

We have also compared our analytic LCs to synthetic ones calculated with  
a one-dimensional radiation hydrodynamics code \verb|STELLA|.
Our analytic LCs are well-reproduced by the numerical modeling.

We have applied our model to only three observed SNe IIn.
We suggest to systematically study SN ejecta and CSM properties of SNe
IIn by applying our LC model to many other SNe IIn.
Such a systematic study will lead to a comprehensive understanding
of SNe IIn, i.e., their progenitors and the mass-loss mechanisms related to them.

\begin{table*}
\centering
\begin{minipage}{50mm}
\caption{List of $\alpha$ from optical and near-infrared LCs.}
\label{table1}
\begin{tabular}{ccc}
\hline
Band&\multicolumn{2}{c}{$\alpha$} \\
&SN 2005ip & SN 2006jd \\
\hline
bolometric &-0.536& -0.0708 \\
$u$ &-1.01 & -0.300 \\
$B$ &-0.923& -0.374\\
$g$ &-0.934& -0.387\\
$V$ &-0.995& -0.451\\
$r$ &-0.854& -0.557\\
$i$ &-1.00& -0.592\\
$Y$ &-0.706& -0.414\\
$J$ &-0.630& -0.137\\
$H$ &-0.171& 0.0950\\
\hline
\end{tabular}
\end{minipage}
\end{table*}

\section*{Acknowledgments}
We thank the anonymous referee for the comments which improved this paper.
T.J.M. and K.M. thank the staff at Stockholm University for their
hospitality during their stay as short-term visitors when this project
was initiated.
T.J.M. is supported by the Japan Society for the Promotion of Science
Research Fellowship for Young Scientists $(23\cdot5929)$.
K.M. acknowledges the financial support by a Grant-in-Aid for
Scientific Research for Young Scientists (23740141).
This research is also supported by World Premier International Research Center Initiative, MEXT, Japan.
The Oskar Klein Centre is funded by the Swedish Research Council.
The work in Russia was supported by RF Government grant 11.G34.31.0047, grants for supporting Scientific Schools 5440.2012.2 and 3205.2012.2, and joint RFBR-JSPS grant 13--02--92119.

\label{lastpage}

\end{document}